# *In-situ* measurements of the radiation stability of amino acids at 15-140 K

Perry A. Gerakines, Reggie L. Hudson, Marla H. Moore, and Jan-Luca Bell

Astrochemistry Laboratory, NASA Goddard Space Flight Center, Greenbelt, MD 20771

## Abstract

We present new kinetics data on the radiolytic destruction of amino acids measured *in situ* with infrared spectroscopy. Samples were irradiated at 15, 100, and 140 K with 0.8-MeV protons, and amino-acid decay was followed at each temperature with and without $H_2O$ present. Observed radiation products included $CO_2$ and amines, consistent with amino-acid decarboxylation. The half-lives of glycine, alanine, and phenylalanine were estimated for various extraterrestrial environments. Infrared spectral changes demonstrated the conversion from the non-zwitterion structure $NH_2$-$CH_2(R)$-$COOH$ at 15 K to the zwitterion structure $^+NH_3$-$CH_2(R)$-$COO^-$ at 140 K for each amino acid studied.

## 1. Introduction

A long-standing hypothesis in the literature is that molecular precursors important for the origin of life on Earth were first formed in space - either in the dense interstellar medium (ISM) from which the Solar System was born or at a later time in the young Solar Nebula - and subsequently delivered to Earth through impact events (Oró 1961; Chyba and Sagan 1992). Each pathway involves low-temperature ices that are exposed to ionizing radiation, both photonic and particle. Indeed, the formation of important bio-molecular species such as amino acids has been documented in laboratory experiments conducted on ices made of molecules known or suspected to be in icy grain mantles (Bernstein et al. 2002; Holtom et al. 2005; Hudson et al. 2008a) and in theoretical models of icy planetary bodies in the early Solar System (Throop 2011). Most studies suggest that these molecules should be present in various space environments.

Astronomers have sought complex organic molecules in deep space, and species as large as $HC_{11}N$ and $C_{70}$ and as complex as ethylene glycol ($HOCH_2CH_2OH$) have been identified in the gas phase of the ISM (see, *e.g.*, Bell et al. 1997; Hollis et al. 2002; Cami et al. 2010). In the Solar System, ethylene glycol has been observed in the comae of comets (Crovisier et al. 2004; Remijan et al. 2008). Unique identifications of such large gas-phase molecules are hindered by the confusion of rotational lines in observed microwave and radio spectra. For example, searches for gas-phase interstellar or cometary glycine have led to inconclusive results (Snyder et al. 2005).

Searches for prebiotic molecules in meteorites and samples returned from comets also have been fruitful. For instance, it has been known for some time that amino acids are present in the organic components of carbonaceous meteorites such as Murchison (Cronin et al. 1979). Recent analyses also have confirmed that many meteoritic amino acids possess an enantiomeric excess and isotopic evidence of formation in low-temperature extraterrestrial environments (Busemann et al. 2006; Glavin et al. 2010). Organics also were detected in the tracks of cometary dust particles collected by the Stardust spacecraft as it flew through the coma of comet 81P/Wild 2 (Sandford et al. 2006). Glycine and amines were among the molecules identified by Elsila et al. (2009).

Since ionizing radiation is an important factor leading to the destruction of molecules in space, it is important to investigate the radiation chemistry and survivability of amino acids and other prebiotic molecules in extraterrestrial environments. Relevant experiments require studying how amino acids respond to various types of radiation at specific temperatures in the solid phase, as opposed to aqueous solutions. Such work has been in progress for many years, going back to at least the 1950s when early electron spin resonance experiments focused on the identification and quantification of free radicals formed from irradiated amino acids in both polycrystalline and single-crystal forms (Combrisson and Uebersfeld 1954; Gordy et al. 1955; Box et al. 1957). Other work concerned reaction products of amino-acid radiolysis, which were identified and quantified at room temperature with both classical chemical analyses, such as volumetric methods, and instrumental techniques, such as gas chromatography (Collinson and Swallow 1956; Meshitsuka et al. 1964; Minegishi et al. 1967). Taken together, these previous studies, and subsequent ones, found that irradiated amino acids decarboxylate and deaminate to produce $CO_2$, amines, $NH_3$, and various free radicals. For a summary of such work, see Sagstuen et al. (2004).

Within the planetary-science and astronomical literature, recent studies of amino-acid survivability include those involving UV photolysis (ten Kate et al. 2006; Orzechowska et al. 2007) and gamma radiolysis (Kminek and Bada 2006). In some cases the amino acids were irradiated at cryogenic temperatures (*e.g.*, 100 K by Orzechowska et al. 2007), but in nearly all cases the chemical and kinetic analyses from such experiments involved room-temperature measurements on samples irradiated and then removed from sealed containers. The far-UV photolytic work of Ehrenfreund et al. (2001) is a rare example of an *in situ* study of amino acid destruction, although even there all experiments were restricted to 12 K.

For planetary and interstellar applications what still are needed are experiments on the radiation-induced destruction of amino acids at multiple temperatures so that trends in the chemistry can be sought. Also, measurements for kinetic analyses should be made *in situ* to avoid the uncertainties introduced by warming samples. Ideally, such work also will make use of MeV protons or keV electrons since they dominate, by number, cosmic radiation, Solar wind particles, and the magnetospheric radiation around Jupiter and Saturn (Moore et al. 2001).

In this paper, we present our first experiments exactly along the lines just described. We report radiation-chemical decomposition data for three amino acids at 15, 100, and 140 K both in the presence and absence of $H_2O$-ice. All measurements were made *in situ* with infrared (IR) spectroscopy from 4000 to 700 cm$^{-1}$ (2.5 to 14 μm). We have calculated half-lives of amino acids expected in the radiation environments of interstellar space and on the surfaces of icy Solar System bodies. Figure 1 shows the chemical structures of the three amino acids we have studied, in both the non-zwitterion and zwitterion forms (see Section 2.3). Unless otherwise noted, the *lævo-* or L isomers of alanine and phenylalanine were used.

## 2. Experimental details

### 2.1. System setup

The experimental system in the Cosmic Ice Laboratory at the NASA Goddard Space Flight Center has been described in detail (Hudson and Moore 1999), and is shown in Fig. 2. It consists of a high-vacuum chamber (P = 5 × 10$^{-7}$ Torr at room temperature) mated to a beam line of a Van de Graaff accelerator and to an IR spectrometer (Nicolet Nexus 670 FT-IR). A polished aluminum substrate (area ≈ 5 cm$^2$) is mounted inside the chamber on the end of the cold finger of a closed-cycle helium cryostat (Air Products Displex DE-204) capable of cooling to a minimum of 15 K. The substrate's temperature is monitored by a silicon diode sensor and can be adjusted up to 300 K using a heater located at the top of the substrate holder. This same substrate is positioned so that the IR spectrometer's beam is reflected from the substrate's surface at a near-normal angle (∼ 5°) and directed onto an HgCdTe (MCT) detector. With an ice sample on the metal substrate, the IR beam passes through the sample before and after reflection. The substrate is fully rotatable through 360° to face the IR spectrometer, the Van de Graaff accelerator, or other components.

For these experiments, a custom-built Knudsen-type sublimation oven was attached to one port of the vacuum chamber, as shown in Fig. 2. The oven consisted of a copper block with a small cavity to hold about 50 mg of amino-acid powder. A copper plate was bolted onto the top of the oven such that a pinhole in the plate was centered over the cavity. The bottom of the copper block was held in contact with a heater consisting of a 100-Ω resistor connected to a 20 V, 225 mA DC power supply. The oven's temperature was monitored by a diode temperature sensor and maintained at a desired set point up to 300°C by a temperature controller. Figure 2 also shows that a deposition tube was positioned such that H$_2$O vapor could be released from it as the substrate faced the amino-acid sublimation oven.

### 2.2. Sample preparation

The compounds, and their purities, used in our experiments were as follows: triply-distilled H$_2$O with a resistivity greater than 10$^7$ Ω cm; glycine, Sigma Chemical Co., 99% purity; D- and L-alanine, Sigma Chemical Co., 98% purity; D- and L-phenylalanine, Sigma Chemical Co., 98% purity.

To prepare a sample, about 50 mg of an amino acid was loaded into the oven cavity, which then was attached to the vacuum chamber, and which then was evacuated overnight to about 5 × 10$^{-7}$ Torr. While facing the IR spectrometer, the substrate was cooled to the desired temperature for sample deposition, and the sublimation oven was heated over ∼ 15 minutes to a set-point temperature of 190°C.  This temperature was sufficient for an adequate sublimation rate for all three amino acids used (Svec and Clyde 1965), but was too low to cause their decomposition.  When 190°C was reached the substrate was turned to face the oven's pinhole opening, which was about 5 cm away.  A baffle positioned between the oven and the substrate prevented contamination of the rest of the vacuum system by the subliming amino acid.

Sample thicknesses were measured during film growth by monitoring the interference fringes of a 650-nm laser's beam reflected from the sample and substrate surfaces with an incidence angle of 10°.  The laser and the detector were mounted inside the sample chamber at the end of the baffle (Fig. 2).  For an oven temperature of 190°C, the rates of film growth varied according to the amino acid from 0.6 to 3.0 $\mu$m hr$^{-1}$.  Final film thicknesses were 0.5-2.0 $\mu$m, depending on the desired sample, below the stopping range for 0.8-MeV protons and ensuring that the entire ice sample was processed.

To create a $H_2O$ + amino acid ice, $H_2O$ vapor was released into the vacuum chamber during the amino acid sublimation by using the deposition tube 1-2 cm in front of the substrate and a metered leak valve.  The valve was calibrated by making deposits at a wide range of valve settings and deriving a general trend of $H_2O$ film growth rate versus valve setting.  This calibration curve and the measured amino-acid growth rates were then used to determine the setting needed to deposit $H_2O$ at a rate that would not obscure the amino-acid IR absorptions.  Based on the chosen growth rates for $H_2O$ and the known growth rates of the amino acids from the sublimation oven, the molecular ratios in the resulting ice mixtures were calculated to be $H_2O$:glycine = 8.7, $H_2O$:alanine = 11, and $H_2O$:phenylalanine = 26.  These calculations assumed mass densities for $H_2O$, glycine, alanine, and phenylalanine of 1.0, 1.61, 1.42, and 1.29 g cm$^{-3}$ respectively, and refractive indices at 650 nm for $H_2O$, glycine, alanine, and phenylalanine of 1.31, 1.46, 1.4, and 1.6 respectively (values from Weast et al. 1984).  Under the experimental conditions described above, some condensation of residual gases in the vacuum system is expected within the sample's volume (on or below the 1 % level) and on the sample's surface.  Neither are expected to have a significant effect on the radiation chemistry, the calculated irradiation doses, or the measured IR spectra over the course of a single experiment, typically lasting 2-4 hours.

*2.3. Amino-acid structures*

Complicating the study of the radiation chemistry of amino acids is that such molecules can be found in multiple forms.  In both the crystalline form supplied commercially and in neutral aqueous solutions (pH ~ 7) the amino acids we studied exist primarily in their so-called zwitterionic forms, $^{+}NH_3$-CH(R)-COO$^{-}$, with a positive charge on the nitrogen and a negative formal charge on an oxygen.  The general formula $H_2N$-CH(R)-COOH for an amino acid corresponds to the non-zwitterionic structure found in the gas phase (Linder et al. 2008) and in matrix-isolated samples (Grenie et al. 1970).  Figure 3 shows glycine as an example, with the zwitterion form being the central structure.

Figure 3 also shows two other glycine structures, the one on the left for acidic conditions ("low" pH) and the one on the right for alkaline conditions ("high" pH).  By working with only pure amino acids and $H_2O$-ice we largely limited each amino acid studied to just two of the possible four structures, namely the zwitterion and non-zwitterion forms of Fig. 1. In all radiation experiments we selected conditions that minimized the abundance of the non-zwitterion form of the amino acid in the sample.

From the roughly twenty amino acids of terrestrial biology we selected three, glycine, alanine, and phenylalanine, for our first study of amino-acid radiolytic stability. The selection of glycine and alanine was in recognition of their simplicity and likely high abundance relative to other extraterrestrial amino acids. The inclusion of phenylalanine was for testing the relative stability conferred, or not, on an irradiated amino acid by an aromatic ring. The "R" in the general amino-acid formula $H_2N-CH(R)-COOH$ is termed a side chain, and in the present paper all three side chains are of low polarity. Going from glycine to alanine in Fig. 1 involves the replacement of a single hydrogen atom with a methyl group ($-CH_3$), while the change from alanine to phenylalanine involves replacing a hydrogen atom with a benzene ring.

Some samples were deposited at 15 K and then heated to 140 K at 2-5 K $min^{-1}$ while recording their IR spectra. This was done prior to irradiation experiments so as to document spectral changes in amino acids due solely to temperature. As described below (in Section 3.1), all samples showed a complete (or nearly complete), irreversible conversion from the non-zwitterion to the zwitterion. Based on this observation, all samples to be irradiated were deposited at 140 K so as to maximize the amount of the zwitterionic form of the amino acid in each case. After a sample's deposition, its temperature was then set to the desired value for the subsequent irradiation.

## 2.4. Radiation doses

All irradiations involved a beam of 0.8 MeV protons at a current of $\sim 1 \times 10^{-7}$ A. For each irradiation, the ice first was rotated to face the proton beam from the accelerator and then turned 180° in the opposite direction after the irradiation so as to face the IR beam of the spectrometer for scanning. See Fig. 2. The dose absorbed by an irradiated amino acid is

$$\text{Dose} \left[ \text{eV molec}^{-1} \right] = mSF \qquad (1)$$

where $m$ is the mass of one amino acid molecule (in g), $S$ is the proton stopping power (in eV $cm^2$ $g^{-1}$ $p+^{-1}$) in that amino acid, and $F$ is the incident proton fluence (in p+ $cm^{-2}$). For $H_2O$ + amino acid mixtures, the energy absorbed by the sample is found when $m$ in equation (1) is replaced with the average mass per molecule, $m_{avg}$. In the astronomical literature such doses often are scaled to 16 amu to provide a basis for comparison among experiments.

The stopping power of 0.8-MeV protons in each sample was calculated using the SRIM software package (Ziegler et al. 2010). For each single-component sample, the ice's density was assumed to be the same as that of the room-temperature crystalline amino acid. For the $H_2O$ + amino acid mixtures, an average mass density was assumed in each case, based on the mixing ratios given in Section 2.2. The resulting stopping powers are listed in Table 1.

Radiation doses may be expressed in a variety of units, and in this work we will use fluence (p+ $cm^{-2}$), eV, eV per molecule, or eV per 16 amu. Doses in units of Mrad are obtained from the fluence $F$ by the expression $SF \cdot 1.60 \times 10^{-20}$ Mrad g eV$^{-1}$, while those in the SI units of MGy are equal to $SF \cdot 1.60 \times 10^{-22}$ MGy g eV$^{-1}$.

## 3. Results

### 3.1. Infrared spectra of unirradiated amino acids during warm-up

Figure 4 shows the mid-IR spectra for the amino acids studied, and Fig. 5 contains the spectra of the $H_2O$ + amino acid mixtures. The 4000-2000 cm$^{-1}$ range includes the C-H stretching region near 3000-2800 cm$^{-1}$ for the acids and the very strong O-H stretching feature near 3200 cm$^{-1}$ due to $H_2O$. The 2000-700 cm$^{-1}$ range is expanded for each spectrum in Figs. 4 and 5 to better show the more-characteristic amino-acid features, such as those of COO$^-$ and NH$_4^+$ groups. Also in each panel of Figs. 4 and 5, the bottom-most curve is the spectrum recorded after sample deposition at the lowest temperature (15 K), and the top three curves, from bottom to top, are the spectra recorded after heating the ice to 50, 100, and 140 K at 2 K min$^{-1}$. Tables 2, 3, and 4 list positions and assignments of the IR absorptions at 15 and 140 K in the samples containing glycine, alanine, and phenylalanine, respectively.

Figure 4 shows that in each amino acid's IR spectrum at 15 K a strong feature is present near 1719 cm$^{-1}$. This band corresponds to the carbonyl (C=O) stretching vibration in the non-zwitterion -COOH group in each amino acid. Figure 4 also shows that this same feature weakens with increasing temperature to 140 K in each sample. The simultaneous disappearance of this absorption and the growth of the COO$^-$ symmetric-stretching vibration near 1406 cm$^{-1}$ indicate that the deposited amino acids transform under heating from the non-zwitterion form into the zwitterionic form. Similar spectral changes have been reported for amorphous valine and isovaline (Hudson et al. 2009), and for glycine and other amino acids in solids at low temperatures (Rosado et al. 1998; Ramaekers et al. 2004). Other spectral features due to the non-zwitterion amino acids (*e.g.*, 1237 cm$^{-1}$ for glycine) also weaken upon heating, while those due to the zwitterion (*e.g.*, those at 1406, 1322, and 1139 cm$^{-1}$ for glycine) develop and strengthen. All features appearing in the spectra are seen to sharpen and grow with temperature. Overall, our IR spectra agree very well with those already in the literature. See, for example, glycine as measured by Khanna et al. (1966) or Maté et al. (2011) or phenylalanine as measured by Hernández et al. (2010).

Our IR spectra of $H_2O$ + amino acids are shown in Fig. 5. In contrast to the spectra in Fig. 4, sharp C=O stretching features are not visible near 1719 cm$^{-1}$ in the lowest-temperature spectra, since they are obscured by the broad, strong $H_2O$ absorption in that region. As the temperature is raised above 100 K, the broad absorption near 1650 cm$^{-1}$ sharpens and the wing near 1700 cm$^{-1}$ becomes less pronounced. These results resemble the observations by Maté et al. (2011), who measured the spectrum of a $H_2O$ + glycine (200:1) mixture at similar temperatures.

### 3.2. Infrared spectra of irradiated samples

Samples to be irradiated were deposited at 140 K to create zwitterionic forms of the amino acid, in order to minimize any confusion with effects other than amino-acid destruction, such as inter-conversion between the non-zwitterion amino acid and the zwitterion. (This temperature was chosen based on the results presented in Section 3.1.) The ices were then cooled to the temperature at which irradiation would take place: 15, 100, or 140 K.

IR spectra of the irradiated samples are displayed in Figs. 6 and 7, in which the features of solid $CO_2$ near 2340 cm$^{-1}$ ($^{12}CO_2$) and 2275 cm$^{-1}$ ($^{13}CO_2$) appear even at the lowest doses. The fundamental vibration of solid CO appears near 2140 cm$^{-1}$ at high doses, possibly as a $CO_2$ destruction product. A series of absorption features near 3300 cm$^{-1}$ appears in the spectra of all three irradiated amino acids and seems to coincide with the features of amines, known decarboxylation products of these molecules. Amines were not detected, however, in irradiated $H_2O$ + amino acid ices, likely because their IR bands were obscured by the very strong $H_2O$ absorption near 3300 cm$^{-1}$ and elsewhere (see Fig. 7).

To monitor amino-acid destruction, the integrated absorption of the COO$^-$ symmetric stretch near 1400 cm$^{-1}$ was measured after deposition and after each radiation dose. A linear baseline was used under the feature to make the integrations. As an example of the resulting trends measured, Fig. 8 shows the normalized area of the 1408 cm$^{-1}$ absorption of glycine versus proton fluence at 15 K. Similar data sets were obtained for each amino acid at 15, 100, and 140 K, both in the presence and absence of $H_2O$-ice.

### 3.3. Amino-acid destruction kinetics

For our kinetic analysis we assumed the form

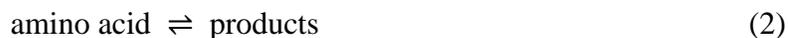

$$\text{amino acid} \rightleftharpoons \text{products} \qquad (2)$$

for amino-acid destruction, with both the forward and reverse reactions taken as first order. This is undoubtedly a simplification, but it proved to be sufficient for fitting the data. Taking glycine (Gly) as an example, the change in its concentration, denoted [Gly], during an irradiation is

$$-\frac{\mathrm{d}[\text{Gly}]}{\mathrm{d}t} = k_1[\text{Gly}] - k_{-1}[\text{products}] \qquad (3)$$

where $k_1$ is the rate constant for the forward reaction and $k_{-1}$ is for the reverse reaction. The usual mathematical treatment (Espenson 1981) gives

$$\ln\frac{[\text{Gly}] - [\text{Gly}]_\infty}{[\text{Gly}]_0 - [\text{Gly}]_\infty} = -(k_1 + k_{-1})t \qquad (4)$$

where 0 and ∞ designate the initial and final amino-acid concentrations, respectively. In practice, we substituted (i) proton fluence $F$ for time t in the above equation and (ii) the integrated absorbance

$$\mathcal{A} = \int Abs(\tilde{v})\, d\tilde{v} \tag{5}$$

of glycine's IR band at 1408 cm$^{-1}$ for [Gly], as the two are proportional. The previous equation then becomes

$$\ln \frac{\mathcal{A} - \mathcal{A}_\infty}{\mathcal{A}_0 - \mathcal{A}_\infty} = -(k_1 + k_{-1})F \tag{6}$$

so that a graph of the left-hand side over fluence $F$ should yield a straight line. The well-known difficulty of finding an accurate value of $\mathcal{A}_\infty$ led us to solve the previous equation for $\mathcal{A}$ and then divide through by $\mathcal{A}_0$ to give

$$\frac{\mathcal{A}}{\mathcal{A}_0} = \left(1 - \frac{\mathcal{A}_\infty}{\mathcal{A}_0}\right) e^{-(k_1 + k_{-1})F} + \frac{\mathcal{A}_\infty}{\mathcal{A}_0} \tag{7}$$

which is of the form $y = a\, e^{-bx} + c$, where $a$, $b$, and $c$ are constants. In particular, parameter $a = (1 - \mathcal{A}_\infty/\mathcal{A}_0)$ is the fractional loss of the amino acid after long times, $b$ is the sum of the rate constants in units of (cm$^2$ p$+^{-1}$), and $c = \mathcal{A}_\infty/\mathcal{A}_0$ is the fraction of each amino acid remaining after prolonged irradiation. Table 5 gives $a$, $b$, and $c$ values found by least-squares curve fits to each experiment's data. Figure 8 shows the curve fit for glycine destruction at 15 K, and Figure 9 shows that the resulting plots from equation (6) were indeed linear for all amino acids.

In most of the radiation experiments reported here, the decay of the zwitterion's COO$^-$ symmetric-stretching feature near 1400 cm$^{-1}$ was well-fit with a function of the form of equation (7). However, for phenylalanine at 140 K, the 1080-cm$^{-1}$ feature was used. In four cases where the final amino-acid concentration was approximately zero, the exponential fit was constrained to $c > -0.05$ without significantly altering its quality.

Radiation-induced changes in a molecule's abundance traditionally are quantified with a so-called $G$ value, defined as the number of molecules altered per 100 eV of energy absorbed. Note that although $G$ is usually called a radiation-chemical yield, it does not correspond to a molecular abundance or a percent yield in the sense in which the term is used by chemists. Instead, the $G$ value of a molecule destroyed or produced by an irradiation is a kinetic quantity, not a thermodynamic one describing an equilibrium situation. We use the notation $G(-M)$ for the $G$ value for the destruction of a molecule, and in this convention their values are always positive.

The $G$ values for amino-acid decay in our experiments were determined from the slopes of the initial, linear behaviors of the exponential fits (Table 5) of the normalized band area vs. $F$ at small doses. Note that as $F \to 0$, the right-hand side of equation (7) is approximately equal to $1 - abF$ (when $a$ and $b$ are substituted as described above), giving an initial slope of $-ab$. The relationship between $-ab$ and the $G$ value of an amino-acid component is given by equation (A.2) in Appendix A, and allows calculation of $G$ in terms of the curve-fit parameters and the sample's properties. Table 6 lists the resulting $G$ value for each amino-acid decay studied, where the uncertainties given are from the propagation of the uncertainties in $a$ and $b$ from Table 5.

As a comparison of the effect of $G$ values on the decay curve of a molecule, Figure 10 contains a plot of the number of glycine molecules versus absorbed energy for the single-component glycine and the $H_2O$+glycine (8.7:1) irradiations at 15 K. Data for the $H_2O$ + glycine mixture have been scaled to match the initial number of glycine molecules in the sample without $H_2O$. In each case, the initial slope of the decay curve indicates the $G$ value for glycine destruction. The glycine sample, which has the steeper initial drop-off, has $G(-M) = 5.8$, whereas the $H_2O$+glycine sample has $G(-M) = 1.9$. It is clear from such a plot that the lifetime of a glycine molecule in an $H_2O$ mixture would be longer than in a sample without $H_2O$.

Some results in the literature indicate that the $G$ values for the decarboxylation of L and D isomers of phenylalanine and leucine by gamma irradiation may differ by factors of about 2-2.5 for doses in the 1-100 krad range (*e.g.*, Merwitz 1976; Tokay et al. 1986). In those studies, the $G$ values of the D and L isomers converge at doses above about 100 krad. In the experiments of the present paper, our minimum fluence of $10^{13}$ p+ cm$^{-2}$ corresponds to about 46,000 krad (0.1 eV molec$^{-1}$). Therefore, we do not expect differences from one enantiomer to the other. As a check, however, we performed proton irradiations of D-alanine and D-phenylalanine at 15 K. To within experimental errors, the resulting values of $G(-M)$ for D-alanine (6.8 ± 0.4) and D-phenylalanine (3.9 ± 0.3) match those for the corresponding L isomers in Table 6.

Figure 11 plots the $G$ values for the cases of the irradiated $H_2O$ + amino acid mixtures versus temperature and amino-acid molar mass. At each temperature, glycine has the highest decay rate, followed by alanine and then phenylalanine. All three amino acids display a monotonic decrease in decay rate with increasing temperature.

# 4. Discussion

## 4.1. Radiation chemistry and destruction rates

Carbon dioxide was the radiation product that was identified with the greatest confidence in our experiments, being seen in all irradiated ices and having a sharp IR absorption unobscured by other features. Although our data are insufficient to elucidate the reaction mechanism, this product is consistent with previous studies performed under different conditions, which have determined that decarboxylation is one of the two primary destruction mechanisms in irradiated amino acids. The details of each mechanism are outlined in a review by Sagstuen et al. (2004). The decarboxylation pathway follows from one-electron oxidation of an amino acid, followed by $H^+$ transfer, and eventually H-atom abstraction from a neighboring molecule to produce an amine and a $CO_2$ molecule as in

$$^+NH_3\text{-}CH(R)\text{-}COO^- \; \rightarrow \; NH_2\text{-}CH_2(R) \; + \; CO_2 \qquad (8)$$

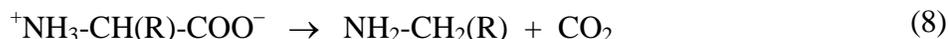

where intermediate steps have been omitted. As before, "R" represents an H atom in the case of glycine, a methyl group ($-CH_3$) in the case of alanine, and the phenylmethyl group ($-CH_2\text{-}C_6H_5$) for phenylalanine. The expected amine products in reaction (8) are methylamine, ethylamine, and phenylethylamine for glycine, alanine, and phenylalanine, respectively. In the cases of the single-component samples, we observed the strong IR feature of $CO_2$ near 2340 cm$^{-1}$ over the course of the irradiation (Fig. 6), and structures in the 3400-3200 cm$^{-1}$ region that could correspond to the absorption features of one or more amines.

The other major amino-acid destruction route presented in previous studies is reductive deamination, which produces ammonia and a free radical:

$$^{+}NH_3\text{-}CH(R)\text{-}COO^{-} \;\; \rightarrow \;\; NH_3 \; + \; \bullet CH(R)COOH \tag{9}$$

where, as before, intermediate steps (outlined by Sagstuen et al. 2004) have been omitted. The radical shown can extract a hydrogen atom to make a carboxylic acid (Sagstuen et al. 2004). Although presumed to be present in our samples, identifying the IR absorption features of the deamination products is problematic due to their positions beneath the strong amino-acid and $H_2O$ absorptions.

The decreasing zwitterion destruction rates with increasing temperature shown in Fig. 11 may have several causes, including  a temperature dependence inherent in the decarboxylation process in solid amino acids (such as the recombination of initially formed radicals). Some evidence may be found in studies of other molecules (such as in the case of $H_2O_2$ studied by Moore and Hudson 2000 or Zheng et al. 2006), where the destruction is countered by a recombination of radicals whose mobility is increased with increasing temperatures.

There is also experimental evidence to support a temperature-dependent inter-conversion between the non-zwitterion and zwitterion forms of the solid amino acids that could affect the net loss of zwitterion at higher temperatures. The non-zwitterion C=O feature near 1720 cm$^{-1}$ is apparent, although weak, in the spectra of samples irradiated at 15 K (see Fig. 6). (The accuracy of this identification is limited by the fact that this weak, broad band is located on the high-wavenumber side of much stronger absorptions near 1600 cm$^{-1}$ from the zwitterion .) However, as indicated in the previous Section and shown in Fig. 4, the non-zwitterion amino acid converts into the zwitterion at 100 and 140 K without irradiation. This thermally-driven process would compete with radiolytic losses of the zwitterion, , thus lowering the observed zwitterion destruction rates. The inter-conversion of zwitterion and non-zwitterion forms would not likely have much of an effect on the observed formation rate of $CO_2$ since both forms of the amino acid decarboxylate when irradiated. Each anhydrous amino acid in Table 6 has its slowest destruction near 100 K. At 140 K, the destruction is slower than that at 15 K and to within the listed uncertainty is similar to that at 100 K. Of the three, alanine has the highest destruction rate at each temperature. Clearer are the trends for the $H_2O$ + amino acid mixtures (Fig. 11), which all show that an increase in temperature from 15 to 140 K  gives a monotonic decrease in the rate of amino acid destruction. Also, of the three amino acids, and at each temperature, glycine is lost the fastest on irradiation in the presence of $H_2O$ while phenylalanine is lost the slowest, which may reflect a kinetic stability conferred by its aromatic side chain.

Kminek and Bada (2006) exposed room-temperature, dry amino acids in sealed glass vials to gamma rays and determined the amino-acid destruction rate constants by chemical analysis. They report destruction rate constants for glycine and alanine of 0.0673 and 0.1127 MGy$^{-1}$ respectively, which correspond to half-life doses (the initial doses required to lower the amino-acid abundance by 50%) of 8.0 and 5.7 eV molec$^{-1}$. As listed in Table 6, we measured 21 eV molec$^{-1}$ for the half-life dose for glycine and 12 eV molec$^{-1}$ for alanine, both at 140 K. The ratio of these half-life doses is in good agreement with those found by Kminek and Bada (2006), although our dose values are each over a factor of two higher. Differences may be due to the fact that different forms of radiation were employed at different temperatures using different experimental conditions and analysis techniques.

Comparing the effects of changing the sample composition, it can be seen from the values listed in Table 6 that the destruction rate ($G$ value) of glycine in a sample containing 8.7 times more $H_2O$ is decreased by factors of 3.1, 2.0, and 1.9 at 15, 100, and 140 K relative to samples of the anhydrous amino acid. Similarly, the $G$ values for alanine destruction are decreased by factors of 4.5, 3.5, 4.9 at 15, 100, and 140 K, respectively, for the $H_2O$ + alanine (11:1) mixtures relative to samples of alanine without $H_2O$. For phenylalanine, the $G$ values in $H_2O$ decrease by 3.0, 2.6, and 5.8 relative to phenylalanine without $H_2O$. These dramatic drops in destruction rates are likely due to the fact that the $H_2O$ molecules may simply shield the amino acids from the incident radiation, without producing additional chemistry, and reduce the amount of energy absorbed per amino-acid molecule. Note that the values of $G$(-M) reported in Table 6 represent the number of molecules lost per unit energy absorbed by the entire ice sample (including the $H_2O$ component).

While such trends in rate data versus temperature or composition are important, it is also desirable to examine equilibrium abundances. Table 5 shows that as $t \rightarrow \infty$ the amino acids showing the larger non-zero concentrations (parameter $c$) in the presence of $H_2O$-ice are glycine and phenylalanine. Overall, this suggests that amino acids with an aromatic side chain are destroyed more slowly and also have a larger final abundance than amino acids with an aliphatic side chain. Work with additional amino acids is needed to test this possibility.

### 4.2. Astrochemical implications

The energy doses required to reduce the initial amino-acid concentration by 50% ("half-life doses") are listed in Table 6 and have been calculated from the decay curves (such as in Fig. 8) for each amino-acid sample. From these doses, we have computed the half-lives of glycine, alanine, and phenylalanine in various extraterrestrial radiation environments, and the results are summarized in Table 7. For the dense ISM, it has been estimated that ices could experience a total cosmic-ray proton dose as high as about 1 eV per 16-amu molecule every $10^6$ yr (Moore et al. 2001; Colangeli et al. 2004). Based on the half-life doses presented in Table 6, the half-lives of glycine, alanine, and phenylalanine in icy interstellar grain mantles would be $\sim 10^7$ yr, roughly the expected lifetime of an interstellar cloud core before gravitational collapse into a protostar. In the cold, diffuse ISM, cosmic-ray fluxes are estimated to be about 10 times higher, resulting in shorter amino-acid lifetimes. Ehrenfreund et al. (2001) predict similar half-lives in the dense ISM due to UV photolysis of amino acid + $H_2O$ ice mixtures, but much shorter half-lives in the diffuse ISM, due to the $\sim 10^5$ times higher UV flux estimated in that environment.

Under the icy surfaces of planetary bodies in the outer Solar System, the energy imparted by bombarding protons is depth-dependent. Oort-cloud comets receive a dose equivalent to 1 eV per 16-amu molecule every $\sim 10^6$ yr at their surfaces and every $\sim 10^8$ yr at 1 cm below their surfaces (Strazzulla et al. 2003). Amino acids in cometary ices would therefore have a half-life of about $10^6$-$10^8$ yr near the surface of a comet. Ices under the surface of Pluto should receive doses of about 1 eV per 16-amu molecule every $5 \times 10^7$ yr at 1 μm depth and after about $1.5 \times 10^8$ yr at 1 m depth (Strazzulla et al. 2003; Hudson et al. 2008b), leading to estimated half-lives of 1-4 $\times 10^8$ yr. On Europa, magnetospheric electrons dominate the radiation environment, with much higher dose rates than in the outer Solar System or interstellar medium (Paranicas et al.

2009). Amino acids on Europa might have a half-life of only a few years at the surface, a few thousand years at 1 cm depth, and longer than about 6-10 million years below 1 m. On Mars, the surface temperature is much higher, about 200 K, and the particle radiation levels are such that a 16-amu molecule would receive a dose of 1 eV every $3\text{-}6 \times 10^7$ yr (Dartnell et al. 2007). On the surface of Mars, an amino acid would have a half-life due to proton bombardment of about $10^8$ yr. Additional effects, such as photolysis by UV photons, may further reduce the expected half-lives.

Given that amino acids have been identified in meteoritic and cometary materials, our results imply that these molecules must have been adequately shielded (by $H_2O$ ice or other materials) from cosmic radiation under the surfaces of meteors and comets after their formation or inclusion in these objects. If amino acids originate in the dense ISM, our results suggest that they could survive long enough for a dense cloud core to collapse into a protostar, where they could become part of the primordial materials out of which comets or planetesimals could then form. In the icy Solar-System environments considered here, most amino acids have half-lives of $\sim$ 10-100 million years at depths greater than a few centimeters. If there is a continuous production of these molecules in radiation environments, recently formed molecules could be found in samples taken at these depths, which are beyond the penetration capability of IR remote sensing techniques, but shallow on the scale of a modern landed exploration mission.

### 5. Summary and conclusions

The infrared spectra of amino acids deposited at 15 K show an irreversible conversion from the non-zwitterion to the zwitterion upon heating to 140 K. Extraterrestrial amino acids will exist in the zwitterionic form if the ices in which they are deposited or created have experienced 140 K or higher temperatures. Experiments performed on the half-lives of amino acids in extraterrestrial environments should therefore take this into account. The radiolytic destruction of amino acids is smaller when $H_2O$-ice is present, and the amount of destruction decreases with increasing temperature. Amino acids have the highest survivability at depths below a few cm in icy planetary bodies within the Solar System.

### Acknowledgments

Zan Peeters constructed and tested the sublimation oven for amino acids. The authors wish to acknowledge support from the NASA Astrobiology Institute (NAI) and the Goddard Center for Astrobiology (GCA), particularly for the summer research internship of JLB. The support of NASA's Exobiology Program is gratefully acknowledged. In addition, we thank Steve Brown, Tom Ward, and Eugene Gerashchenko, members of the Radiation Effects Facility at NASA Goddard Space Flight Center, for operation of the proton accelerator.

## Appendix A.  Derivation of equations for *G* values

The *G* value for the destruction of an amino acid, written $G(-M)$, is the decrease in the number of amino-acid molecules in a sample divided by the number of 100 eV doses absorbed after an exposure to an ion fluence *F*.  The dose absorbed (in eV) is given by $D = \rho h A S F$, where $\rho$ is the sample's density (in g cm$^{-3}$), *h* is its thickness (in cm), *A* is its surface area (in cm$^2$), *S* is the stopping power (in eV cm$^2$ g$^{-1}$ p+$^{-1}$), and *F* is the ion fluence (in p+ cm$^{-2}$).

We determined the number of molecules lost during an irradiation using the area under an IR absorption feature as a function of dose (see Fig. 8).  Since the number of molecules *N* in a thin sample is directly proportional to the area $\mathcal{A}$ under an IR absorption band, the number of amino acid molecules remaining after an irradiation exposure is given by $N = N_0(\mathcal{A}/\mathcal{A}_0)$, where $\mathcal{A}_0$ is the initial band area and $N_0$ is the initial number of amino acid molecules.  For a sample with a ratio of $H_2O$ to amino acid equal to *R* and an average molecular mass of $m_{avg}$ (in g), $N_0 = N_{total}/(R+1)$, where $N_{total}$ is the total number of all molecules in the sample, equal to $\rho h A/m_{avg}$ (note that in a single-component sample $R = 0$).

The *G* value for the destruction of an amino acid in a sample may then be found by taking the ratio of the number of molecules lost $(N_0 - N)$ and the energy absorbed (*D*) and multiplying by 100 to convert from eV$^{-1}$ units to (100 eV)$^{-1}$ units, leading to

$$G(-M) = \frac{100}{(R+1)m_{avg}S} \frac{(1 - \mathcal{A}/\mathcal{A}_0)}{F} \tag{A.1}$$

Since amino-acid decay follows the form of equation (7), then for early irradiation steps (when $F \to 0$) the normalized area is $(\mathcal{A}/\mathcal{A}_0) \approx 1 - abF$, which, when substituted into equation (A.1) leads to a relation between the curve fit parameters of Table 5 and *G*:

$$G(-M) = \frac{100ab}{(R+1)m_{avg}S} \tag{A.2}$$

This formula was used to calculate the values of *G* that are listed in Table 6.

Table 1.  Properties of 0.8 MeV protons and amino-acid ice samples.

| Sample | Stopping power $S$ [eV cm$^2$ g$^{-1}$ p+$^{-1}$] | Average molar mass $M_{avg}$ [g mol$^{-1}$] | Absorbed dose at $F = 10^{14}$ p+ cm$^{-2}$ | |
|---|---|---|---|---|
| | | | [eV molec$^{-1}$] | [eV (16 amu)$^{-1}$] |
| Glycine | $2.8 \times 10^8$ | 75 | 3.5 | 0.74 |
| Alanine | $2.9 \times 10^8$ | 89 | 4.2 | 0.76 |
| Phenylalanine | $2.9 \times 10^8$ | 165 | 7.9 | 0.77 |
| | | | | |
| $H_2O$ + glycine (8.7:1) | $2.9 \times 10^8$ | 23.9 | 1.1 | 0.76 |
| $H_2O$ + alanine (11:1) | $2.9 \times 10^8$ | 23.8 | 1.1 | 0.77 |
| $H_2O$ + phenylalanine (26:1) | $2.9 \times 10^8$ | 23.4 | 1.1 | 0.77 |

Table 2.  Positions (cm$^{-1}$) and assignments of glycine IR absorptions.

| Sample | | | | |
|---|---|---|---|---|
| Glycine 15 K | Glycine 140 K | H$_2$O + Gly 15 K | H$_2$O + Gly 140 K | Assignment(s)* |
| 2912 s | 2911 s | 2927 m | | $\nu$(N-H), $\nu$(C-H) |
| | | 2790 w | 2762 vw | |
| | | 2744 w | | |
| 2640 w | 2646 m | 2655 m | 2663 w | |
| 2546 s | | 2574 s | | |
| 1970 m | 2085 m | | | |
| 1719 s | 1721 w‡ | | | $\nu$(C=O) |
| 1587 s | 1588/1510 s | 1559 w | | $\delta$(NH$_2$), $\nu_a$(COO$^-$) |
| | | | 1532 s | $\delta$(NH$_3$) |
| | 1439 w | 1444 m | 1442 s | $\delta$(CH$_2$) |
| 1408 s† | 1406 s | 1420 m† | 1417 vs | $\nu_s$(COO$^-$) |
| 1324 m | 1322 s | 1333 m | 1330 vs | $\nu$(C-C), $\omega$(CH$_2$), $\omega$(NH$_3$) |
| | | 1310 w | | |
| 1237 s | 1248† m | 1249 s | 1257 m | $\nu$(C-O) + $\delta$(COH) |
| 1172 w | | 1175 m | | $\tau$(CH$_2$) |
| | 1139 m | | 1150 s | $\rho$(NH$_3$) |
| 1119 w | | 1120 w | | $\nu$(C-N) |
| | | 1089 vw | 1091 vw | |
| 1037 m | 1040 m | 1042 m | 1044 m | $\tau$(CO), $\nu$(C-N) |
| 936 m | 930 m | | | $\nu$(C-C) + $\omega$(NH$_2$), $\rho$(CH$_2$) |
| 893 w† | 891 m | | | $\nu$(C-C) |
| 868 w | | | | $\omega$(NH$_2$) + $\nu$(C-C) |
| 670 m† | 672 m | | | $\delta$(COO) |

References: Rosado et al. (1998); Fischer et al. (2005); Maté et al. (2011).
vs = very strong, s = strong, m = medium, w = weak, vw = very weak
*$\nu$ = stretch, $\delta$ = bend, $\omega$ = wag, $\tau$ = torsion, $\rho$ = rock, s = symmetric, a = asymmetric
†Due to glycine zwitterion in original deposit.
‡Due to non-zwitterion form remaining in sample after warm-up.

Table 3.  Positions (cm$^{-1}$) and assignments of alanine IR absorptions.

| Sample | | | | |
|---|---|---|---|---|
| Alanine 15 K | Alanine 140 K | $H_2O$ + Ala 15 K | $H_2O$ + Ala 140 K | Assignment(s) |
| 2980 m | 2982 m | 2982 w | 2983 vw | $\nu_a(CH_3)$, $\nu(NH_3^+)$ |
| 2941 m | 2945 m | 2946 w | 2948 vw | $\nu(CH)$ |
| 2890 m | | 2752 w | 2751 w | |
| | 2620 m | 2623 w | 2651 m | $2\delta(NH_3^+)$ |
| 2537 s | 2534 m | 2555 w | 2550 m | |
| 1980 m | 2126 m | | | |
| 1715 s | 1718 vw | 1709 s | | |
| 1589 vs | 1585 vs | 1595 vs | 1591 vs | $\nu_a(COO^-)$ |
| | 1525 m | 1557 s | 1543 vs | $\delta(NH_3^+)$ |
| 1463 s | 1460 s | 1464 s | 1464 s | $\delta_a(CH_3)$ |
| 1407 s | 1406 vs | 1415 s | 1416 vs | $\nu_s(COO^-)$ |
| 1370 m | 1369 m | 1375 m | 1375 m | $\delta_a(CH_3)$ |
| 1349 m | 1349 m | 1354 m | 1354 s | $\delta_s(CH_3)$ |
| 1301 m | 1301 m | 1306 m | 1304 s | $\delta(CH)$ |
| 1255 s | | 1264 s | | $\rho(NH_3^+)$ |
| 1220 s | 1222 m | 1229 s | 1228 m | |
| 1140 m | 1141 w | 1142 m | 1152 m | $\rho(CH_3)$ |
| 1121 m | | 1119 m | 1116 m | $\rho(NH_3^+)$ |
| 1082 m | | 1086 m | | |
| | 1009 w | 1015 m | 1014 w | $\nu_s(CCNC)$ |
| 977 m | | 985 m | | $2\tau(NH_3)$ |
| 914 m | 916 w | 919 w | 920 w | $\nu_a(CCNC)$, $\rho(NH_3^+)$ |
| 843 m | 843 m | 847 w | 847 w | $\nu(NC)$, $\rho(CH3)$ |
| 819 m | 823 w | 821 w | | |
| 771 w | | 777 w | 776 w | $\delta(COO^-)$ |
| 746 w | | 752 vw | | $\delta(COO^-)$ |
| | 668 w | | | |

References: Wang and Storms (1971); Susi and Byler (1980); Rozenberg et al. (2003); Hernández et al. (2009).  Notations are as defined in Table 2.

Table 4.  Positions (cm$^{-1}$) and assignments of phenylalanine IR absorptions.

| Sample | | | | |
|---|---|---|---|---|
| Phe 15 K | Phe 140 K | H$_2$O + Phe 15 K | H$_2$O + Phe 140 K | Assignment(s) |
| 3359 m | 3355 w | | | |
| 3268 s | 3259 m | | | |
| 3164 w | | | | |
| 3109 w | 3107 vw | | | |
| 3088 m | 3088 w | | | |
| 3064 s | 3065 m | | | |
| 3030 s | 3030 s | | | |
| 3005 m | 3005 m | | | |
| 2928 s | 2927 s | | | |
| 2859 m | 2859 m | | | |
| 2508 s | 2566 s | 2543 m | 2681 w | |
| 1960 m | 1959 w | 1967 vw | 1965 vw | |
| 1895 w | 1898 w | 1896 vw | | |
| 1717 vs | 1715 m | | | |
| 1629 vs | 1625 vs | 1648 vs | | $\delta(NH_3^+)$ |
| 1605 m | 1605 w | 1607 w | 1607 m | $\nu_a(COO^-)$, $\nu(C\text{-}C)_{ring}$ |
| 1585 w | 1584 vw | 1548 w | 1545 s | $\nu_a(COO^-)$ |
| | 1515 m | | | |
| 1497 vs | 1497 vs | 1499 m | 1499 m | $\delta(NH_3^+)$ |
| 1455 s | 1455 s | 1457 m | 1457 m | $\nu(C\text{-}C)_{ring}$ |
| | | | 1446 m | |
| 1397 s | 1399 s | 1412 m | 1413 s | $\nu_s(COO^-)$ |
| | | 1359 vw | 1359 m | |
| 1333 m | 1334 s | 1343 m | 1341 m | $\delta(CH_2)$ |
| | | 1323 vw | | |
| 1282 w | 1289 vw | | 1288 vw | |
| 1238 m | 1240 w | 1246 m | 1247 w | $\delta(CH_2)$ |
| 1209 w | 1211 w | 1208 w | 1213 w | |
| 1157 w | 1154 m | 1155 w | 1157 w | $\delta(CH)_{ring}$ |
| 1113 w | 1112 vw | 1117 vw | | |
| 1079 m | 1081 m | 1082 m | 1084 w | $\delta(CH)_{ring}$ |
| | | | 1046 w | |
| 1031 m | 1031 w | 1031 m | 1032 w | $\delta(CH)_{ring}$ |
| 1003 w | 1003 vw | | | $\delta(C\text{-}C)_{ring}$ |
| 960 w | 970 vw | | | |
| 919 m | 914 m | | | $\delta(CH)_{ring}$ |
| 904 m | | | | |
| 858 m | 857 m | | | $\delta(CH)_{ring}$ |
| 819 m | 814 w | | | |
| 794 w | 794 vw | | | |
| 754 m | 747 m | 752 m | | |
| 700 m | 700 m | 703 m | 702 w | |

References: Olsztynska et al. (2001); Hernández et al. (2010).  Notations are as defined in Table 2.

Table 5.  Parameters for amino-acid radiolytic decay.

| Sample | T [K] | Curve fit parameters* | | |
|--------|-------|------|------|------|
| | | $a$ | $b$ [cm$^2$ p+$^{-1}$] | $c$ |
| Glycine | 15 | 0.91 ± 0.02 | $(2.2 ± 0.2) \times 10^{-15}$ | 0.09 ± 0.02 |
| | 100 | 1.1 ± 0.1 | $(1.2 ± 0.1) \times 10^{-15}$ | -0.05 ± 0.1† |
| | 140 | 0.96 ± 0.06 | $(1.2 ± 0.2) \times 10^{-15}$ | 0.01 ± 0.07 |
| Alanine | 12 | 0.93 ± 0.01 | $(3.3 ± 0.1) \times 10^{-15}$ | 0.07 ± 0.01 |
| | 100 | 1.0 ± 0.1 | $(1.9 ± 0.1) \times 10^{-15}$ | -0.03 ± 0.02 |
| | 140 | 1.1 ± 0.1 | $(2.1 ± 0.2) \times 10^{-15}$ | -0.05 ± 0.05† |
| Phenylalanine | 15 | 0.85 ± 0.03 | $(2.8 ± 0.3) \times 10^{-15}$ | 0.12 ± 0.03 |
| | 100 | 1.1 ± 0.1 | $(1.5 ± 0.3) \times 10^{-15}$ | -0.05 ± 0.1† |
| | 140 ‡ | 0.73 ± 0.02 | $(3.6 ± 0.2) \times 10^{-15}$ | 0.27 ± 0.02 |
| H$_2$O + Gly (8.7:1) | 15 | 0.71 ± 0.04 | $(2.9 ± 0.4) \times 10^{-15}$ | 0.26 ± 0.04 |
| | 100 | 0.79 ± 0.06 | $(2.5 ± 0.3) \times 10^{-15}$ | 0.18 ± 0.06 |
| | 140 | 0.87 ± 0.03 | $(2.1 ± 0.1) \times 10^{-15}$ | 0.12 ± 0.03 |
| H$_2$O + Ala (11:1) | 15 | 0.98 ± 0.01 | $(2.2 ± 0.1) \times 10^{-15}$ | 0.03 ± 0.01 |
| | 100 | 1.0 ± 0.1 | $(1.8 ± 0.3) \times 10^{-15}$ | -0.05 ± 0.08† |
| | 140 | 1.0 ± 0.1 | $(1.5 ± 0.1) \times 10^{-15}$ | -0.01 ± 0.02 |
| H$_2$O + Phe (26:1) | 15 | 0.79 ± 0.03 | $(3.9 ± 0.4) \times 10^{-15}$ | 0.17 ± 0.02 |
| | 100 | 0.80 ± 0.02 | $(3.1 ± 0.3) \times 10^{-15}$ | 0.16 ± 0.03 |
| | 140 | 0.87 ± 0.03 | $(2.0 ± 0.2) \times 10^{-15}$ | 0.17 ± 0.04 |

*All fits have the functional form $ae^{-bF} + c$.

† Fit constrained to $c \geq -0.05$.

‡ For phenylalanine at 140 K, the 1080-cm$^{-1}$ feature was used in the fit.  See text for details.

Table 6.  Amino-acid $G$ values and half-life doses.

| Sample | T [K] | $G$(−M)† [molec (100 eV)$^{-1}$] | Half-life dose‡ | |
| | | | [eV molec$^{-1}$] | [eV (16 amu)$^{-1}$] |
| --- | --- | --- | --- | --- |
| Glycine | 15 | 5.8 ± 0.5 | 13 ± 1.4 | 2.7 ± 0.3 |
| | 100 | 3.6 ± 0.4 | 19 ± 3.7 | 4.1 ± 1.0 |
| | 140 | 3.3 ± 0.6 | 21 ± 5.7 | 4.5 ± 1.2 |
| Alanine | 15 | 7.2 ± 0.2 | 9.9 ± 0.4 | 1.8 ± 0.1 |
| | 100 | 4.6 ± 0.1 | 15 ± 1.0 | 2.7 ± 0.2 |
| | 140 | 5.4 ± 0.6 | 12 ± 2.3 | 2.2 ± 0.4 |
| Phenylalanine | 15 | 3.0 ± 0.3 | 25 ± 3.6 | 2.4 ± 0.4 |
| | 100 | 2.1 ± 0.5 | 32 ± 12 | 3.1 ± 1.2 |
| | 140 | 3.3 ± 0.2 | 25 ± 2.5 | 2.5 ± 0.2 |
| $H_2O$ + Gly (8.7:1) | 15 | 1.9 ± 0.3 | 4.8 ± 1.0 | 3.2 ± 0.6 |
| | 100 | 1.8 ± 0.3 | 4.5 ± 1.1 | 3.0 ± 0.7 |
| | 140 | 1.7 ± 0.1 | 4.6 ± 0.5 | 3.1 ± 0.3 |
| $H_2O$ + Ala (11:1) | 15 | 1.6 ± 0.1 | 3.6 ± 0.1 | 2.4 ± 0.1 |
| | 100 | 1.3 ± 0.2 | 4.2 ± 1.2 | 2.8 ± 0.8 |
| | 140 | 1.1 ± 0.1 | 5.0 ± 0.4 | 3.4 ± 0.2 |
| $H_2O$ + Phe (26:1) | 15 | 1.0 ± 0.1 | 2.9 ± 0.4 | 2.0 ± 0.2 |
| | 100 | 0.81 ± 0.08 | 3.6 ± 0.5 | 2.4 ± 0.2 |
| | 140 | 0.57 ± 0.06 | 4.8 ± 0.9 | 3.3 ± 0.6 |

† Number of amino-acid molecules destroyed per 100 eV absorbed by the sample.

‡ Energy dose required to reduce the initial amino-acid abundance by 50%.

Table 7.  Energy doses and estimated amino-acid half-lives.

| Location | Temp [K] | Depth [cm] | Radiation Dose Rate [eV $(16\text{ amu})^{-1}\text{ yr}^{-1}$] | Half-Life[a] [yr] | | | | | |
| --- | --- | --- | --- | --- | --- | --- | --- | --- | --- |
| | | | | Gly | Ala | Phe | Gly in $H_2O$ | Ala in $H_2O$ | Phe in $H_2O$ |
| Mars[b] | 200 | 0 | $2.9\times10^{-8}$ | $1.9\times10^{8}$ | $9.6\times10^{7}$ | $9.7\times10^{7}$ | $9.6\times10^{7}$ | $1.2\times10^{8}$ | $1.1\times10^{8}$ |
| | | 100 | $1.7\times10^{-8}$ | $3.3\times10^{8}$ | $1.7\times10^{8}$ | $1.7\times10^{8}$ | $1.7\times10^{8}$ | $2.1\times10^{8}$ | $2.0\times10^{8}$ |
| Europa[c] | 100 | $10^{-3}$ | 1.8 | 2.3 | 1.5 | 1.7 | 1.6 | 1.5 | 1.2 |
| | | 1 | $2.2\times10^{-3}$ | 1,800 | 1,200 | 1,400 | 1,300 | 1,200 | 990 |
| | | 100 | $3.6\times10^{-7}$ | $1.1\times10^{7}$ | $7.6\times10^{6}$ | $8.7\times10^{6}$ | $8.0\times10^{6}$ | $7.5\times10^{6}$ | $6.2\times10^{6}$ |
| Pluto[d] | 40 | $10^{-4}$ | $2.2\times10^{-8}$ | $1.4\times10^{8}$ | $9.3\times10^{7}$ | $1.2\times10^{8}$ | $1.4\times10^{8}$ | $1.1\times10^{8}$ | $8.8\times10^{7}$ |
| | | 100 | $6.5\times10^{-9}$ | $4.7\times10^{8}$ | $3.1\times10^{8}$ | $4.0\times10^{8}$ | $4.5\times10^{8}$ | $3.7\times10^{8}$ | $2.9\times10^{8}$ |
| Comets/Outer Sol. Sys.[e] | 40 | $10^{-4}$ | $1.0\times10^{-6}$ | $3.1\times10^{6}$ | $2.0\times10^{6}$ | $2.6\times10^{6}$ | $2.9\times10^{6}$ | $2.4\times10^{6}$ | $1.9\times10^{6}$ |
| | | 1 | $7.7\times10^{-9}$ | $4.0\times10^{8}$ | $2.6\times10^{8}$ | $3.4\times10^{8}$ | $3.8\times10^{8}$ | $3.1\times10^{8}$ | $2.5\times10^{8}$ |
| Cold Diffuse ISM[f] | 40 | -- | $3.2\times10^{-6}$ | $9.6\times10^{5}$ | $6.3\times10^{5}$ | $8.1\times10^{5}$ | $9.2\times10^{5}$ | $7.5\times10^{5}$ | $6.0\times10^{5}$ |
| Dense ISM[g] | 10 | -- | $1.6\times10^{-7}$ | $1.7\times10^{7}$ | $1.1\times10^{7}$ | $1.4\times10^{7}$ | $1.8\times10^{7}$ | $1.6\times10^{7}$ | $1.1\times10^{7}$ |

[a] The total time to reduce initial amino-acid abundance by 50%, as determined by the parameters given in Table 5 for each sample and the radiation dose rates for each listed environment.  Decay rates at 40 and 200 K have been determined by interpolation and extrapolation of the measured rates.

[b] Dartnell et al. (2007).

[c] Total dose rate includes electrons; see Paranicas et al. (2009).  Equitorial averaged temperature of 100 K; see Spencer et al. (1999).

[d] Hudson et al. (2008b).

[e] Data for a heliocentric distance of 85 AU from Strazzulla et al. (2003).

[f] Moore et al. (2001).

[g] Moore et al. (2001); Colangeli et al. (2004).

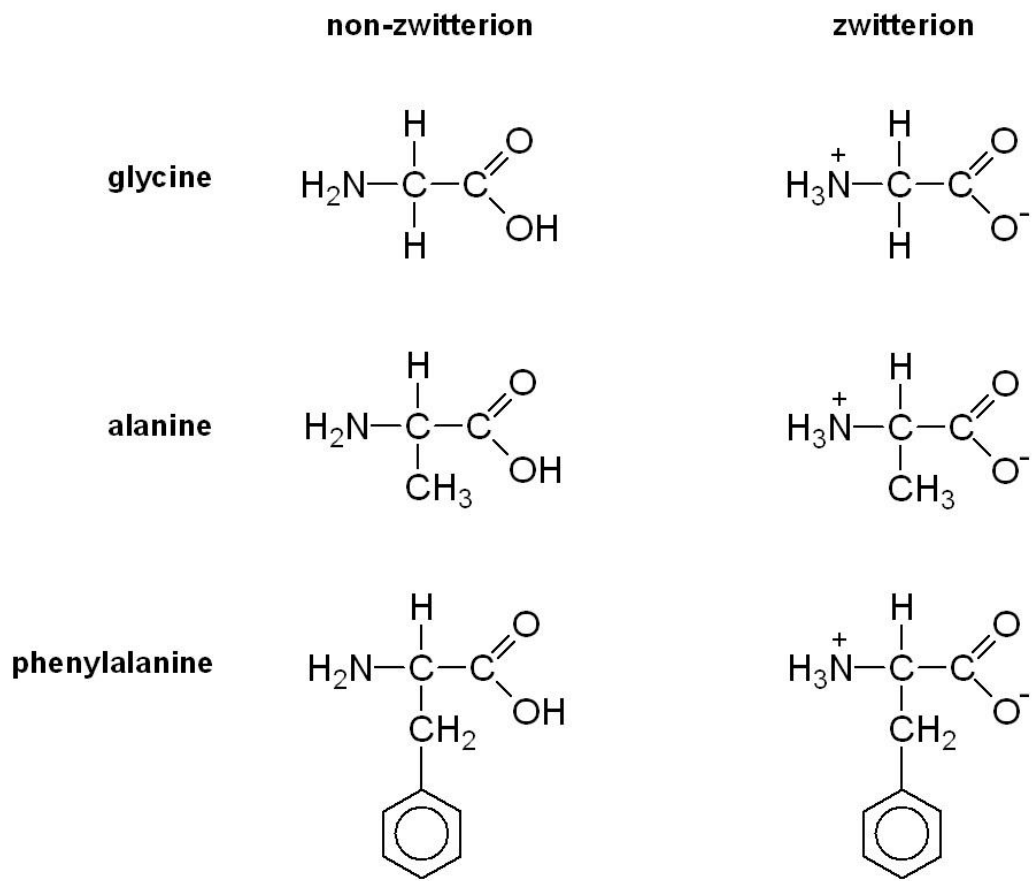

**Figure 1.** Chemical structures of the amino acids studied.

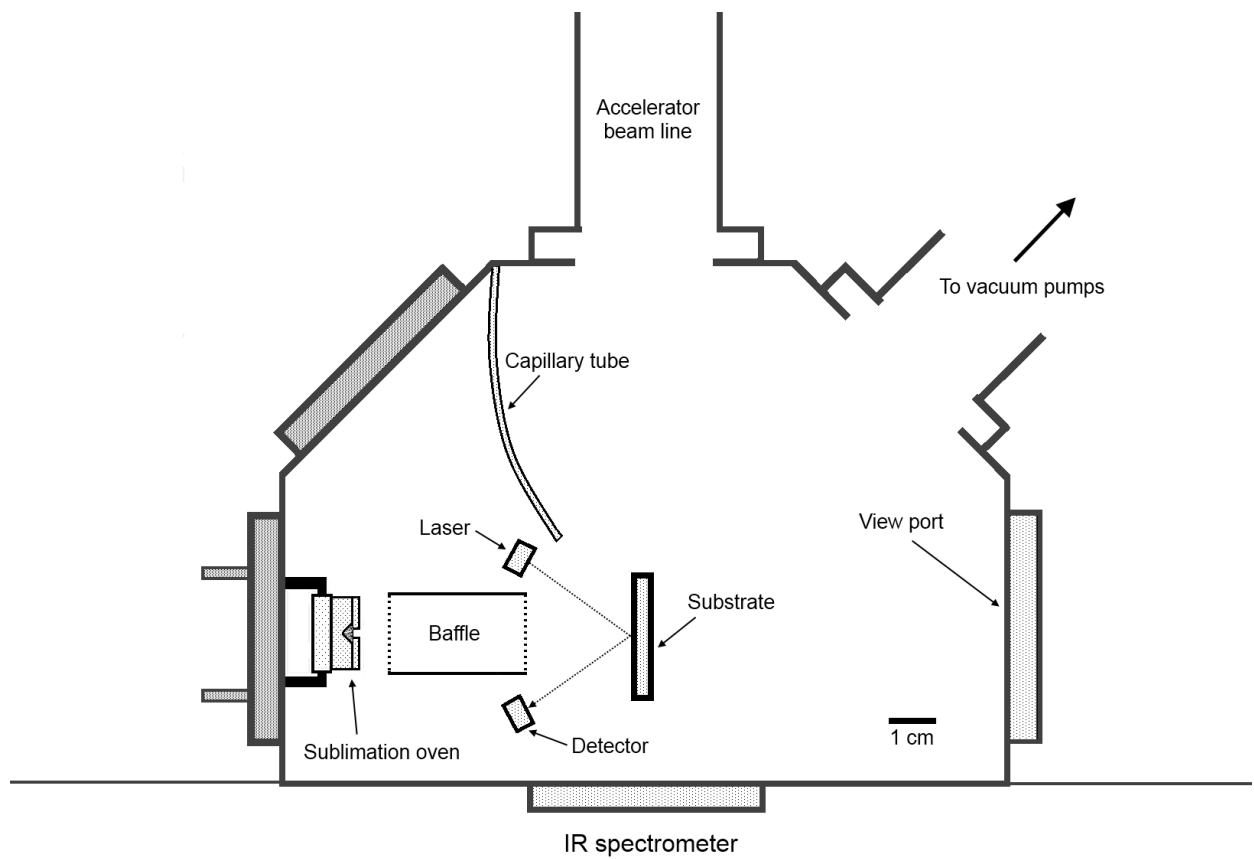

**Figure 2.** Schematic of the experimental set-up used to create ice samples with amino acids. Amino-acid powders evaporate from a sublimation oven, pass through a baffle (a 2-cm diameter copper tube), and collect on the cold substrate. To create a mixture, a gas is simultaneously released into the vacuum in front of the substrate through a capillary tube. Film growth is monitored using the interference fringes from a 650-nm laser reflected from the sample and substrate surfaces. The substrate may be rotated to face the oven (as shown), the IR spectrometer beam, or a beam of 0.8-MeV protons from a Van de Graaff accelerator. Scale is approximate.

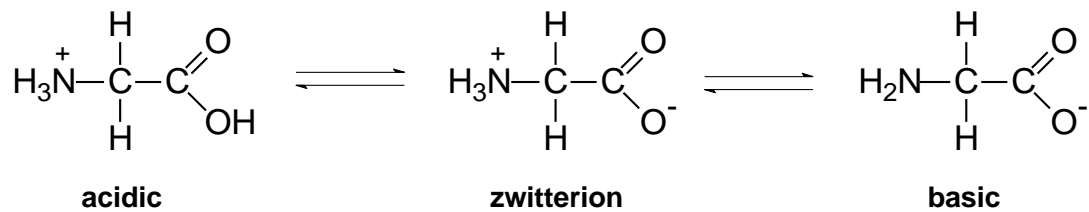

**Figure 3.** The forms of glycine in acidic, neutral, and alkaline solutions.

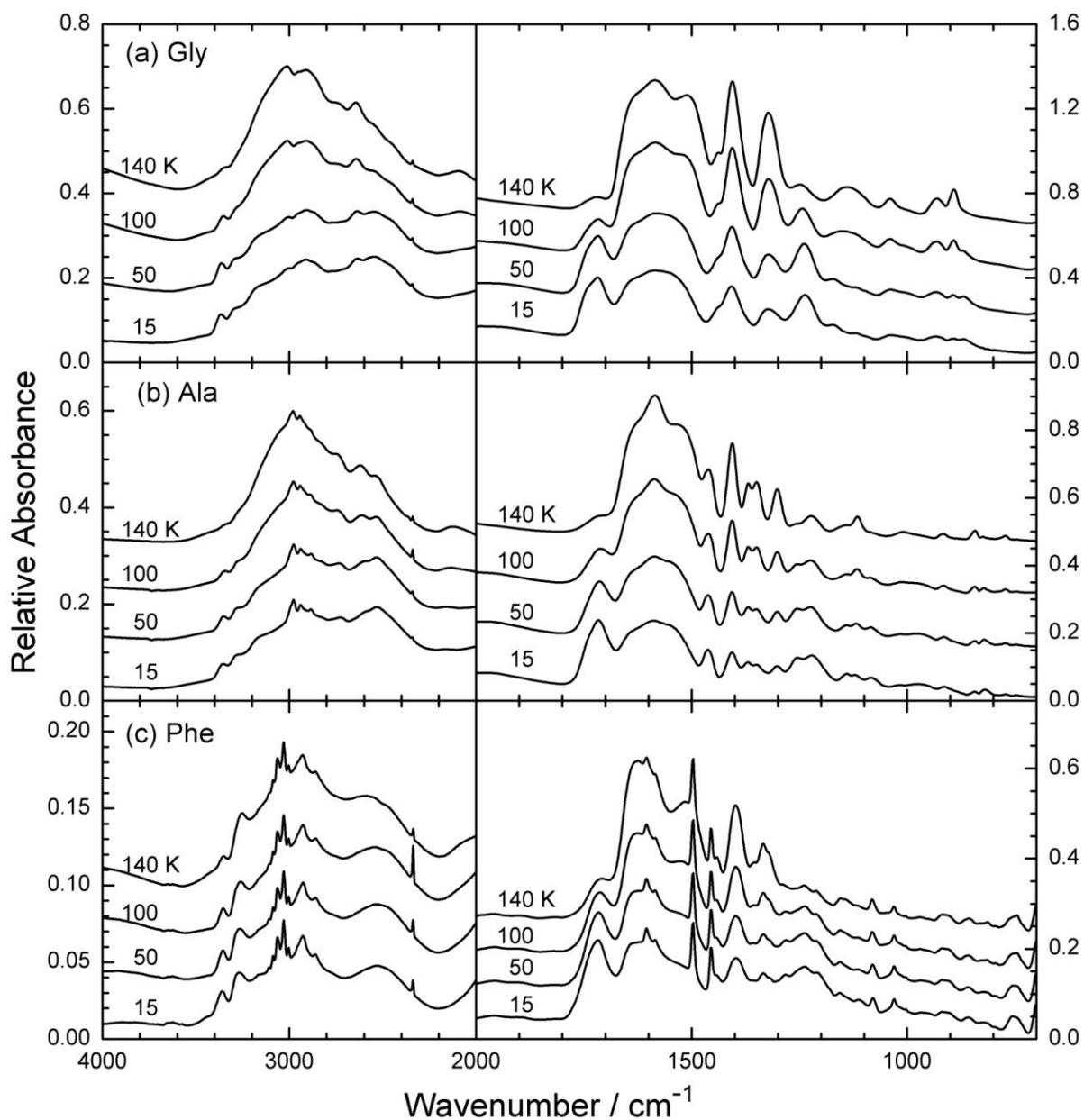

**Figure 4.** Infrared spectra of amino-acid samples after deposition at 15 K and heating to 50, 100, and 140 K: (a) glycine, (b) alanine, (c) phenylalanine. The left and right sides of panels (a)-(c) have different vertical scales, as indicated on the axes. Spectra have been offset for clarity.

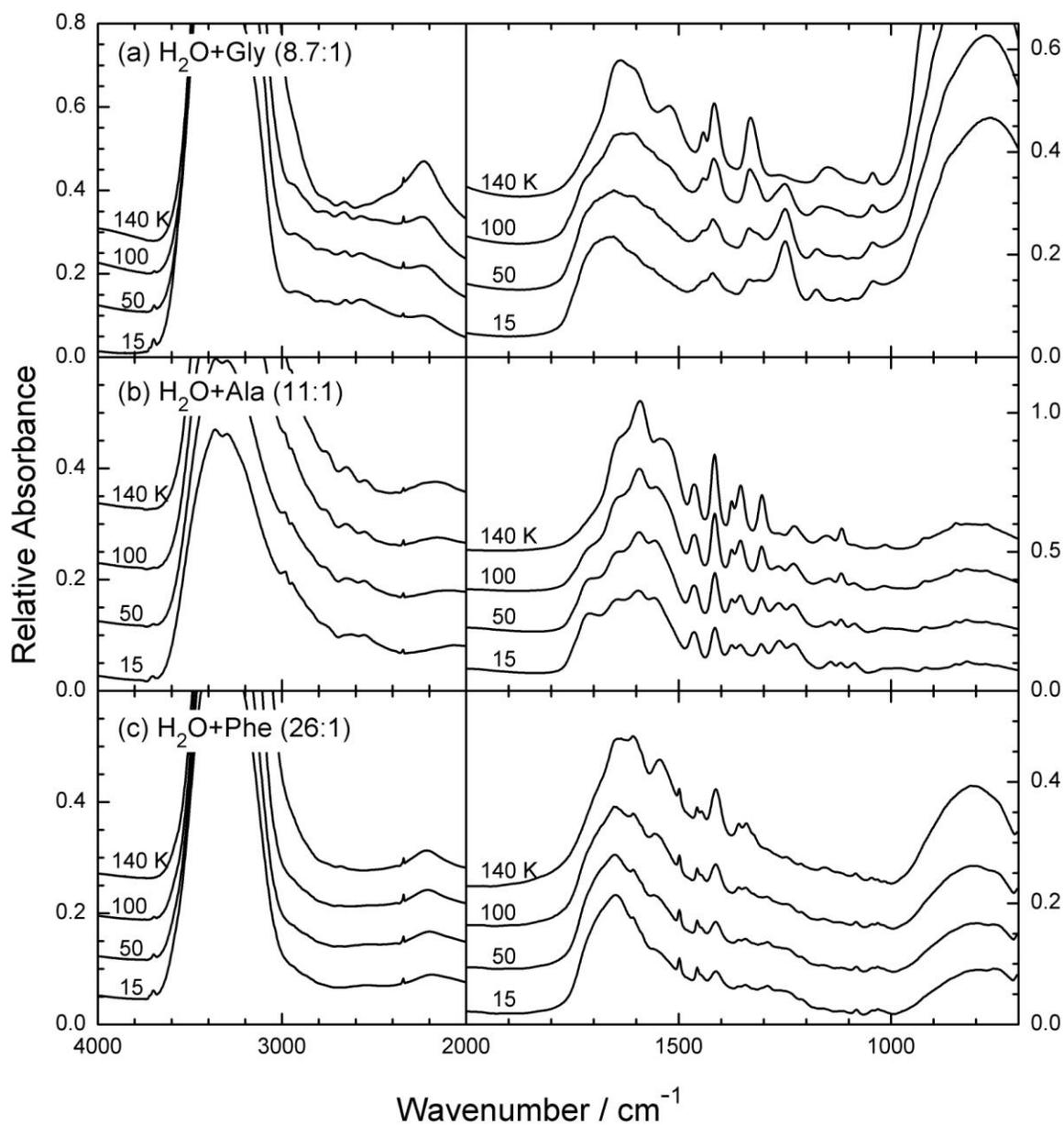

**Figure 5.** Infrared spectra of $H_2O$ + amino acid ices after deposition at 15 K and heating to 50, 100, and 140 K: (a) $H_2O$ + glycine (8.7:1), (b) $H_2O$ + alanine (11:1), (c) $H_2O$ + phenylalanine (26:1). The left and right sides of panels (a)-(c) have different vertical scales, as indicated on the axes. Spectra have been offset for clarity.

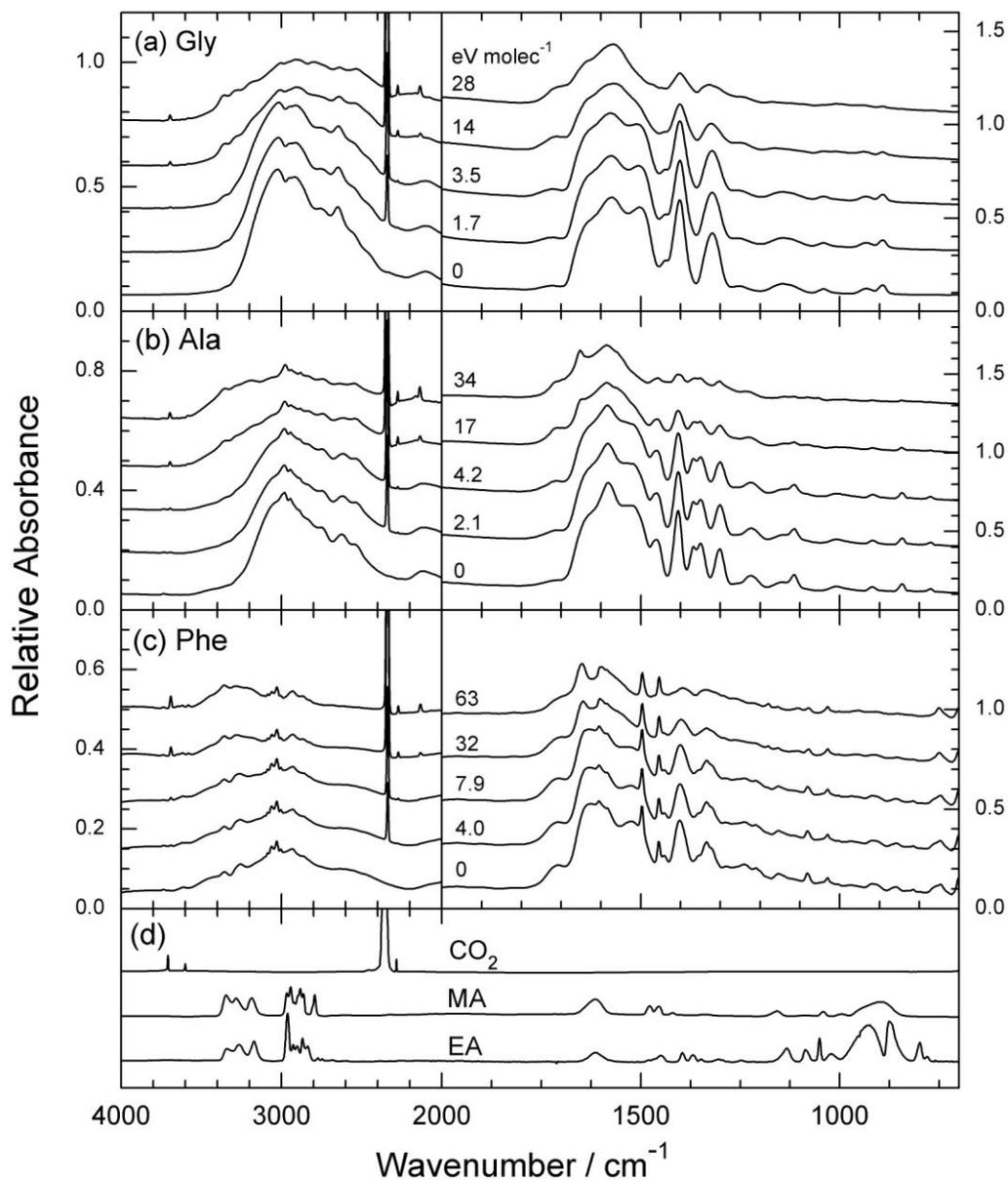

**Figure 6.** Infrared absorbance spectra during irradiation at 15 K of (a) glycine, (b) alanine, and (c) phenylalanine. In each case spectra are shown after fluences of 0, $5.0 \times 10^{13}$, $1.0 \times 10^{14}$, $4.0 \times 10^{14}$, and $8.0 \times 10^{14}$ p+ cm$^{-2}$ (from bottom to top, labeled with corresponding values of eV per amino-acid molecule). The left and right sides of panels (a)-(c) have different vertical scales, as indicated on the axes. Spectra have been offset for clarity. The bottom panel (d) contains 15-K reference spectra of $CO_2$, methylamine (MA, $CH_3NH_2$), and ethylamine (EA, $CH_3CH_2NH_2$) ices.

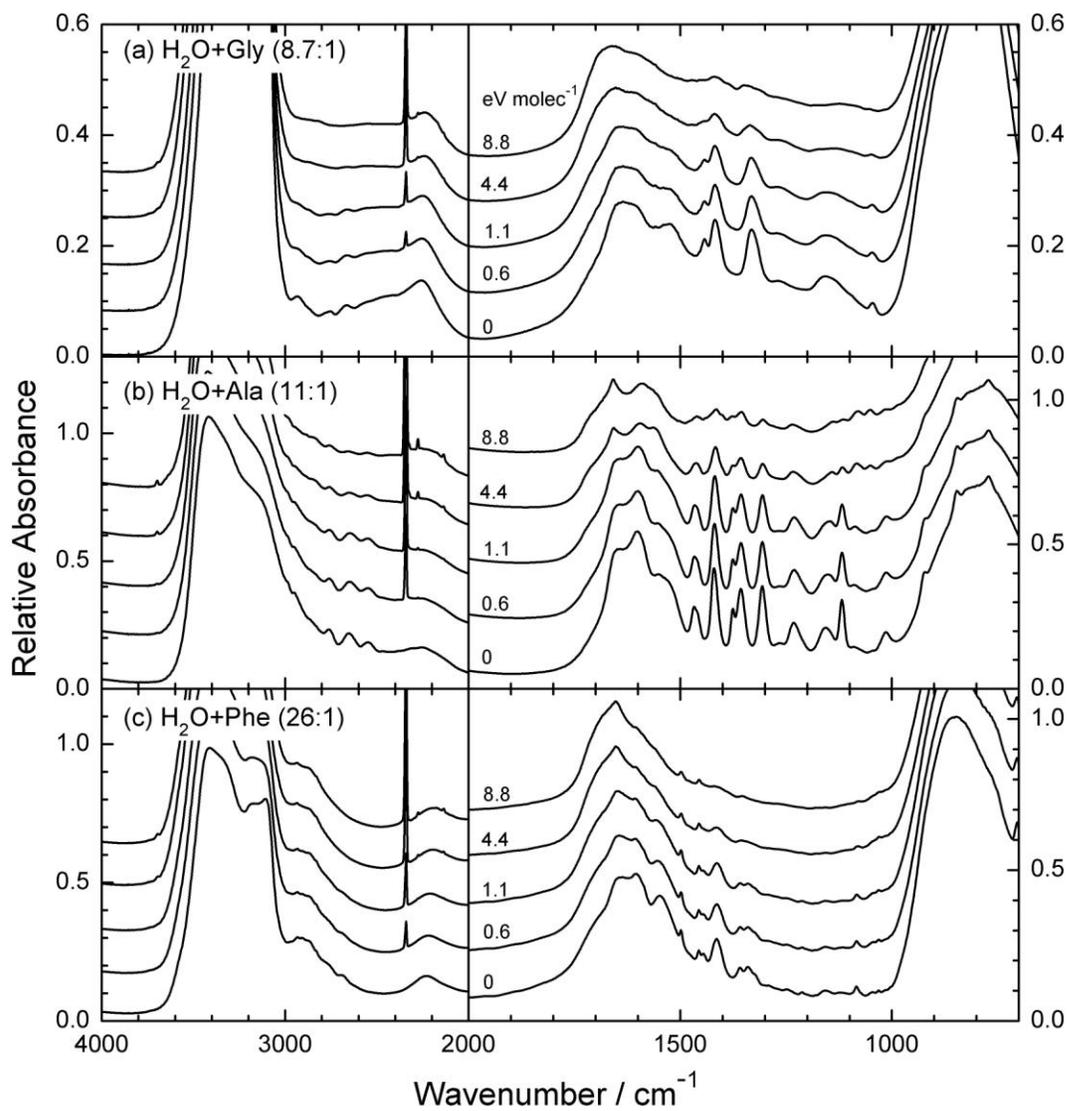

**Figure 7.** Infrared spectra during irradiation at 15 K of (a) $H_2O$ + glycine (8.7:1), (b) $H_2O$ + alanine (11:1), and (c) $H_2O$ + phenylalanine (26:1). In each case spectra are shown after fluences of 0, $5.0 \times 10^{13}$, $1.0 \times 10^{14}$, $4.0 \times 10^{14}$, and $8.0 \times 10^{14}$ p+ cm$^{-2}$ (from bottom to top, labeled with corresponding values of eV per molecule). The left and right sides of panels (a)-(c) have different vertical scales, as indicated on the axes. Spectra have been offset for clarity.

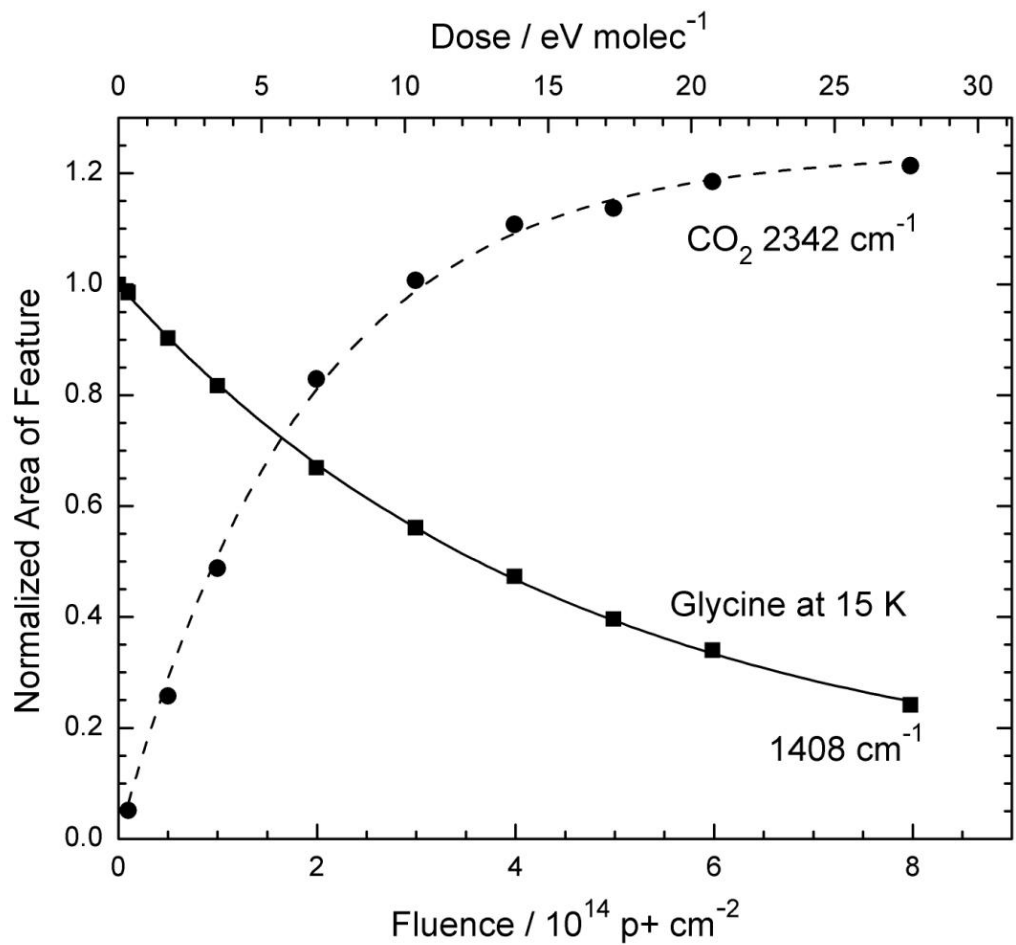

**Figure 8.** Areas of the 1408 cm$^{-1}$ feature of glycine and the 2342 cm$^{-1}$ feature of $CO_2$ during the irradiation of a glycine sample at 15 K, each normalized to the initial area of the 1408 cm$^{-1}$ feature. The solid curve is a fit of the form of equation (7) to the glycine data. The dashed curve is a fit to the $CO_2$ data of the form $y = p(1 - e^{-qF})$.

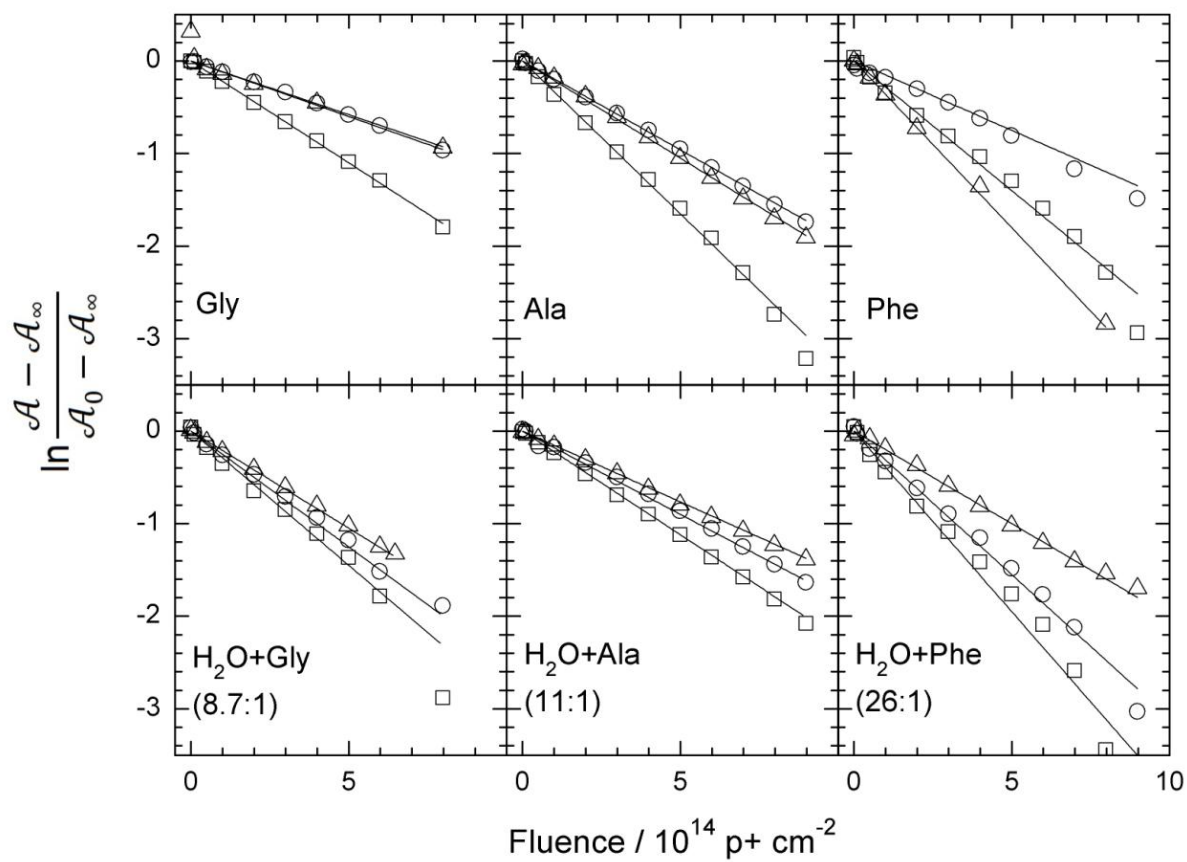

**Figure 9.** Exponential decays of the zwitterion features for ices at 15, 100, and 140 K. In each case, the curve fit parameters from Table 5 have been used to plot the left-hand side of equation (6) versus fluence. Square symbols represent data for irradiation at 15 K, circles 100 K, and triangles 140 K. The lines in each case have a slope equal to the corresponding value of $-b$ from Table 5.

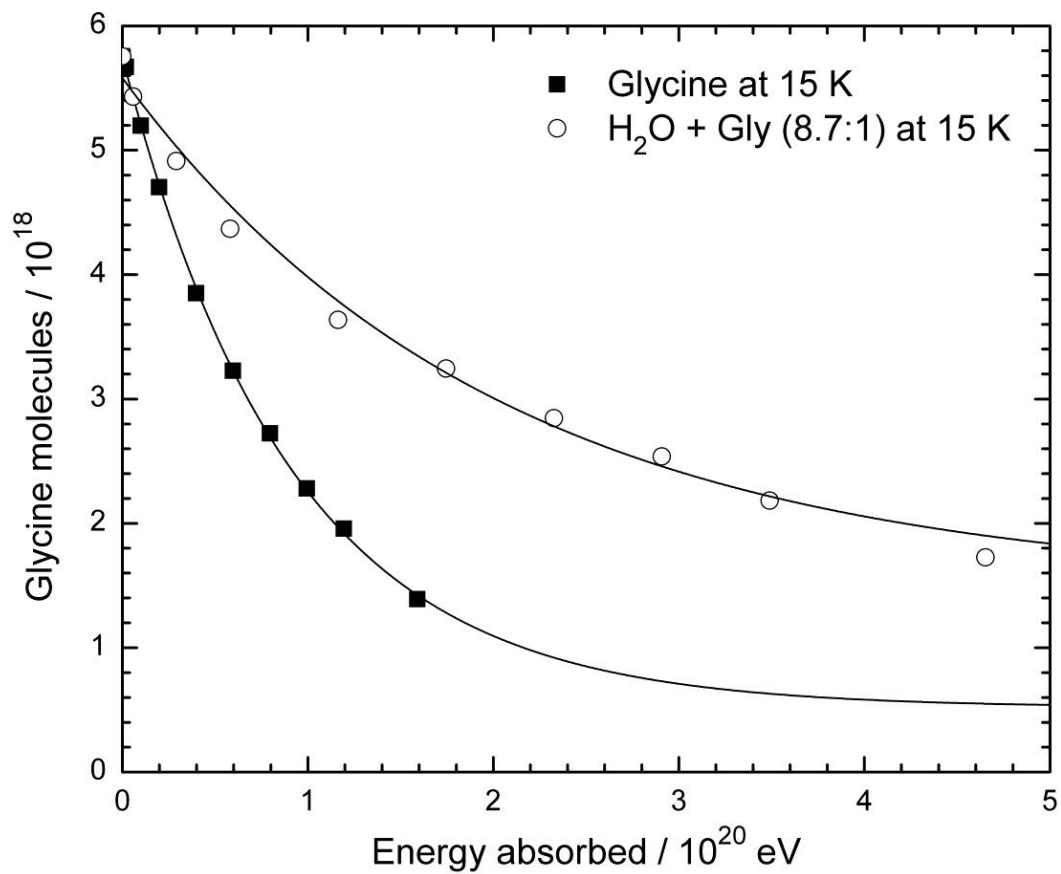

**Figure 10.** Comparison of the decays of the glycine and $H_2O$ + glycine samples at 15 K. Data for the $H_2O$ mixture have been scaled such that the graph represents samples with the same initial number of glycine molecules. Solid lines are the fits from Table 5 scaled to the appropriate units.

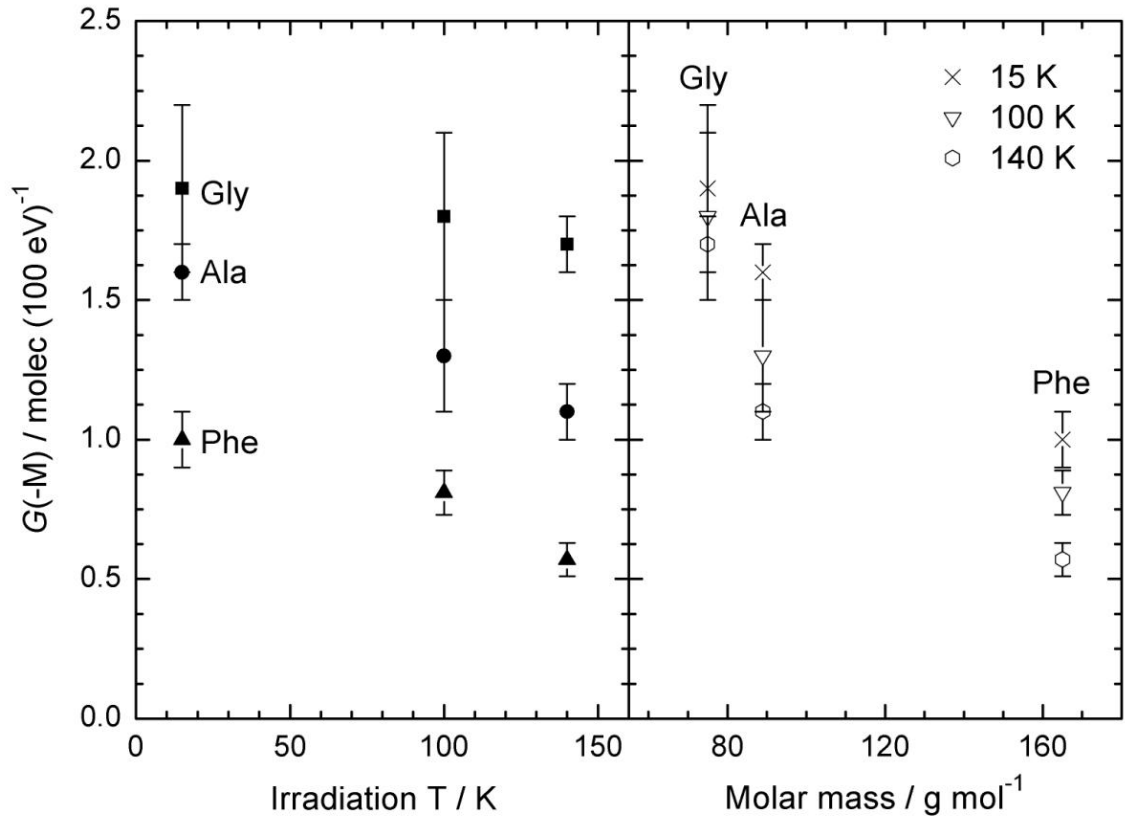

**Figure 11.** Amino acid destruction in $H_2O$ mixtures– left panel: $G$(-M) vs. irradiation temperature; right panel: $G$(-M) vs. vs. amino-acid molar mass.

**REFERENCES**

Bell, M. B., Feldman, P. A., Travers, M. J., McCarthy, M. C., Gottlieb, C. A., Thaddeus, P., 1997. Detection of $HC_{11}N$ in the cold dust cloud TMC-1. Astrophys. J. 483, L61-L64.

Bernstein, M. P., Dworkin, J. P., Sandford, S. A., Cooper, G. W., Allamandola, L. J., 2002. Racemic amino acids from the ultraviolet photolysis of interstellar ice analogues. Nature 416, 401-403.

Box, H. C., Freund, H., Hoffman, J. G., 1957. Paramagnetic resonance studies of radiation effects in powder amino acids. Radiat. Res. 7, 305-305.

Busemann, H., Young, A. F., Alexander, C. M. O., Hoppe, P., Mukhopadhyay, S., Nittler, L. R., 2006. Interstellar chemistry recorded in organic matter from primitive meteorites. Science 312, 727-730.

Cami, J., Bernard-Salas, J., Peeters, E., Malek, S. E., 2010. Detection of $C_{60}$ and $C_{70}$ in a young planetary nebula. Science 329, 1180-1182.

Chyba, C., Sagan, C., 1992. Endogenous production, exogenous delivery, and impact-shock synthesis of organic molecules - an inventory for the origins of life. Nature 355, 125-132.

Colangeli, L., Brucato, J. R., Bar-Nun, A., Hudson, R. L., Moore, M. H., 2004. Laboratory experiments on cometary materials. in: M. C. Festou, H. U. Keller, H. A. Weaver, (Eds.), Comets II. University of Arizona Press, Tucson, pp. 695-717.

Collinson, E., Swallow, A. J., 1956. The radiation chemistry of organic substances. Chem. Rev. 56, 471-568.

Combrisson, J., Uebersfeld, J., 1954. Detection de la resonance paramagnetique dans certaines substances organiques irradiees. Cr. Hebd. Acad. Sci. 238, 1397-1398.

Cronin, J. R., Pizzarello, S., Moore, C. B., 1979. Amino acids in an antarctic carbonaceous chondrite. Science 206, 335-337.

Crovisier, J., Bockelee-Morvan, D., Biver, N., Colom, P., Despois, D., Lis, D. C., 2004. Ethylene glycol in comet C/1995 O1 (Hale-Bopp). Astron. Astrophys. 418, L35-L38.

Dartnell, L. R., Desorgher, L., Ward, J. M., Coates, A. J., 2007. Modelling the surface and subsurface Martian radiation environment: Implications for astrobiology. Geophys. Res. Lett. 34, L0227.

Ehrenfreund, P., Bernstein, M. P., Dworkin, J. P., Sandford, S. A., Allamandola, L. J., 2001. The photostability of amino acids in space. Astrophys. J. 550, L95-L99.

Elsila, J. E., Glavin, D. P., Dworkin, J. P., 2009. Cometary glycine detected in samples returned by Stardust. Meteorit. Planet. Sci. 44, 1323-1330.

Espenson, J. H., 1981. Chemical Kinetics and Reaction Mechanisms. McGraw-Hill, New York.

Fischer, G., Cao, X. L., Cox, N., Francis, M., 2005. The FT-IR spectra of glycine and glycylglycine zwitterions isolated in alkali halide matrices. Chem. Phys. 313, 39-49.

Glavin, D. P., Aubrey, A. D., Callahan, M. P., Dworkin, J. P., Elsila, J. E., Parker, E. T., Bada, J. L., Jenniskens, P., Shaddad, M. H., 2010. Extraterrestrial amino acids in the Almahata Sitta meteorite. Meteorit. Planet. Sci. 45, 1695-1709.

Gordy, W., Ard, W. B., Shields, H., 1955. Microwave spectroscopy of biological substances .1. Paramagnetic resonance in X-irradiated amino acids and proteins. Proc. Nat. Acad. Sci. 41, 983-996.

Grenie, Y., Lassegues, J.-C., Garrigou-Lagrange, C., 1970. Infrared spectrum of matrix-isolated glycine. J. Chem. Phys. 53, 2980-2982.

Hernández, B., Pfluger, F., Adenier, A., Kruglik, S. G., Ghomi, M., 2010. Vibrational analysis of amino acids and short peptides in hydrated media. VIII. Amino acids with aromatic side chains: L-phenylalanine, L-tyrosine, and L-tryptophan. J. Phys. Chem. B 114, 15319-15330.

Hernández, B., Pflüger, F., Nsangou, M., Ghomi, M., 2009. Vibrational analysis of amino acids and short peptides in hydrated media. IV. Amino acids with hydrophobic side chains: L-alanine, L-valine, and L-isoleucine. J. Phys. Chem. B 113, 3169-3178.

Hollis, J. M., Lovas, F. J., Jewell, P. R., Coudert, L. H., 2002. Interstellar antifreeze: ethylene glycol. Astrophys. J. 571, L59-L62.

Holtom, P.D., Bennett, C.J., Osamura, Y., Mason, N.J., Kaiser, R.I., 2005. A combined experimental and theoretical study on the formation of the amino acid glycine ($NH_2CH_2COOH$) and its isomer ($CH_3NHCOOH$) in extraterrestrial ices. Astrophys. J. 626, 940-952.

Hudson, R. L., Lewis, A. S., Moore, M. H., Dworkin, J. P., Martin, M. P., 2009. Enigmatic isovaline: investigating the stability, racemization, and formation of a non-biological meteoritic amino acid. Bioastronomy 2007: Molecules, Microbes, and Extraterrestrial Life 420, 157-162.

Hudson, R. L., Moore, M. H., 1999. Laboratory studies of the formation of methanol and other organic molecules by water+carbon monoxide radiolysis: Relevance to comets, icy satellites, and interstellar ices. Icarus 140, 451-461.

Hudson, R. L., Moore, M. H., Dworkin, J. P., Martin, M. P., Pozun, Z. D., 2008a. Amino acids from ion-irradiated nitrile-containing ices. Astrobiology 8, 771-779.

Hudson, R. L., Palumbo, M. E., Strazzulla, G., Moore, M. H., Cooper, J. F., Sturner, S. J., 2008b. Laboratory studies of the chemistry of trans-Neptunian object surface materials. in: M. A. Barucci, H. Boehnhardt, D. P. Cruikshank, A. Morbidelli, (Eds.), The Solar System Beyond Neptune. University of Arizona Press, Tucson, pp. 507-523.

Khanna, R. K., Horak, M., Lippincott, E. R., 1966. Infrared studies on glycine and its addition compounds. Spectrochim. Acta 22, 1759-1771.

Kminek, G., Bada, J. L., 2006. The effect of ionizing radiation on the preservation of amino acids on Mars. Earth Planet. Sci. Lett. 245, 1-5.

Linder, R., Seefeld, K., Vavra, A., Kleinermanns, K., 2008. Gas phase infrared spectra of nonaromatic amino acids. Chem. Phys. Lett. 453, 1-6.

Maté, B., Rodriguez-Lazcano, Y., Gálvez, Ó., Tanarro, I., Escribano, R., 2011. An infrared study of solid glycine in environments of astrophysical relevance. Phys. Chem. Chem. Phys. 13, 12268-12276.

Merwitz, O., 1976. Selective Radiolysis of Enantiomers. Radiat. Environ. Biophys. 13, 63-69.

Meshitsuka, G., Shindo, K., Minegishi, A., Suguro, H., Shinozaki, Y., 1964. Radiolysis of solid glycine. Bull. Chem. Soc. Jpn. 37, 928-930.

Minegishi, A., Shinozaki, Y., Meshitsuka, G., 1967. Radiolysis of solid L-□-alanine. Bull. Chem. Soc. Jpn. 40, 1271-1272.

Moore, M. H., Hudson, R. L., 2000. IR detection of $H_2O_2$ at 80 K in ion-irradiated laboratory ices relevant to Europa. Icarus 145, 282-288.

Moore, M. H., Hudson, R. L., Gerakines, P. A., 2001. Mid- and far-infrared spectroscopic studies of the influence of temperature, ultraviolet photolysis and ion irradiation on cosmic-type ices. Spectrochim. Acta A 57, 843-858.

Olsztynska, S., Komorowska, M., Vrielynck, L., Dupuy, N., 2001. Vibrational spectroscopic study of L-phenylalanine: Effect of pH. Appl. Spectr. 55, 901-907.

Oró, J., 1961. Comets and formation of biochemical compounds on primitive Earth. Nature 190, 389-390.

Orzechowska, G. E., Goguen, J. D., Johnson, P. V., Tsapin, A., Kanik, I., 2007. Ultraviolet photolysis of amino acids in a 100 K water ice matrix: application to the outer Solar System bodies. Icarus 187, 584-591.

Paranicas, C., Cooper, J. F., Garrett, H. B., Johnson, R. E., Sturner, S. J., 2009. Europa's radiation environment and its effects on the surface. in: R. T. Pappalardo, W. B. McKinnon, K. Khurana, (Eds.), Europa. University of Arizona Press, Tucson, pp. 529-544.

Ramaekers, R., Pajak, J., Lambie, B., Maes, G., 2004. Neutral and zwitterionic glycine-H2O complexes: a theoretical and matrix-isolation Fourier transform infrared study. J. Chem. Phys. 120, 4182-4193.

Remijan, A. J., Milam, S. N., Womack, M., Apponi, A. J., Ziurys, L. M., Wyckoff, S., A'Hearn, M. F., de Pater, I., Forster, J. R., Friedel, D. N., Palmer, P., Snyder, L. E., Veal, J. M., Woodney, L. M., Wright, M. C. H., 2008. The distribution, excitation, and formation of cometary molecules: methanol, methyl cyanide, and ethylene glycol. Astrophys. J. 689, 613-621.

Rosado, M. T., Duarte, M. L. T. S., Fausto, R., 1998. Vibrational spectra of acid and alkaline glycine salts. Vib. Spectrosc. 16, 35-54.

Rozenberg, M., Shoham, G., I, R., Fausto, R., 2003. Low-temperature Fourier transform infrared spectra and hydrogen bonding in polycrystalline L-alanine. Spectrochim Acta A 59, 3253-3266.

Sagstuen, E., Sanderud, A., Hole, E. O., 2004. The solid-state radiation chemistry of simple amino acids, revisited. Radiat. Res. 162, 112-119.

Sandford, S. A., et al., 2006. Organics captured from comet 81P/Wild 2 by the Stardust spacecraft. Science 314, 1720-1724.

Snyder, L. E., Lovas, F. J., Hollis, J. M., Friedel, D. N., Jewell, P. R., Remijan, A., Ilyushin, V. V., Alekseev, E. A., Dyubko, S. F., 2005. A rigorous attempt to verify interstellar glycine. Astrophys. J. 619, 914-930.

Spencer, J. R., Tamppari, L. K., Martin, T. Z., Travis, L. D., 1999. Temperatures on Europa from Galileo Photopolarimeter-Radiometer: Nighttime Thermal Anomalies. Science 284, 1514.

Strazzulla, G., Cooper, J. F., Christian, E. R., Johnson, R. E., 2003. Ion irradiation of TNOs: from the fluxes measured in space to the laboratory experiments. C. R. Phys. 4, 791-801.

Susi, H., Byler, D. M., 1980. Vibrational analysis of L-alanine and deuterated analogs. J. Mol. Struct. 63, 1-11.

Svec, H. J, Clyde, D. D., 1965. Vapor pressures of some α-amino acids. J. Chem. Eng. Data 10, 151-152.

ten Kate, I. L., Garry, J. R. C., Peeters, Z., Foing, B., Ehrenfreund, P., 2006. The effects of Martian near-surface conditions on the photochemistry of amino acids. Planet. Space Sci. 54, 296-302.

Throop, H. B., 2011. UV photolysis, organic molecules in young disks, and the origin of meteoritic amino acids. Icarus 212, 885-895.

Tokay, R. K., Norden, B., Liljenzin, J. O., 1986. Has nuclear chirality been a prebiotic source of optical purity of living systems - the quantum yields of gamma-decarboxylation and beta-decarboxylation of 1-C-14 labeled D-leucine and L-leucine in the solid-state can indicate considerable selectivity. Orig. Life Evol. Biosph. 16, 421-422.

Wang, C. H., Storms, R. D., 1971. Temperature-dependent Raman study and molecular motion in L-alanine single crystal. J. Chem. Phys. 55, 3291-3299.

Weast, R. C., Astle, M. J., Beyer, W. H. (Eds.), 1984. The CRC Handbook of Chemistry and Physics, 64th Edition.  CRC Press, Boca Raton.

Zheng, W., Jewitt, D., Kaiser, K. I., 2006.  Temperature dependence of the formation of hydrogen, oxygen, and hydrogen peroxide in electron-irradiated crystalline water ice. Astrophys. J. 648, 753-761.

Ziegler, J. F., Ziegler, M. D., Biersack, J. P., 2010. SRIM - The stopping and range of ions in matter. Nucl. Instrum. Methods Phys. Res., Sect. B 268, 1818-1823.